\documentclass[journal]{IEEEtran}

\usepackage{graphicx}
\usepackage{algorithmic}
\usepackage{algorithm}
\usepackage{multirow}
\usepackage{cite}
\usepackage{subfigure}
\usepackage{amsthm}
\usepackage{amssymb}
\usepackage{amsfonts}
\usepackage{amsmath}

\begin{document}

\title{Achieving Small World Properties using Bio-Inspired Techniques in Wireless Networks}

\author{\IEEEauthorblockN{Rachit Agarwal\IEEEauthorrefmark{1}, Abhik Banerjee\IEEEauthorrefmark{1}\IEEEauthorrefmark{2}, Vincent Gauthier\IEEEauthorrefmark{1}, Monique Becker\IEEEauthorrefmark{1}, Chai Kiat Yeo\IEEEauthorrefmark{2} and Bu Sung Lee\IEEEauthorrefmark{2}}
\IEEEauthorblockA{\IEEEauthorrefmark{1}Lab. CNRS SAMOVAR UMR 5157, Telecom Sud Paris, Evry, France}\\
\IEEEauthorblockA{Email: {\{rachit.agarwal, vincent.gauthier, monique.becker\}}@telecom-sudparis.eu}\\
\IEEEauthorblockA{\IEEEauthorrefmark{2} CeMNet, School of Computer Engineering, Nanyang Technological University, Singapore}\\
\IEEEauthorblockA{Email: \{abhi0018, asckyeo, ebslee\}@ntu.edu.sg}}

\maketitle


\begin{abstract}
It is highly desirable and challenging for a wireless ad hoc network to have self-organization properties in order to achieve network wide characteristics. Studies have shown that Small World properties, primarily low average path length and high clustering coefficient, are desired properties for networks in general. However, due to the spatial nature of the wireless networks, achieving small world properties remains highly challenging. Studies also show that, wireless ad hoc networks with small world properties show a degree distribution that lies between geometric and power law. In this paper, we show that in a wireless ad hoc network with non-uniform node density with only local information, we can significantly reduce the average path length and retain the clustering coefficient. To achieve our goal, our algorithm first identifies logical regions using Lateral Inhibition technique, then identifies the nodes that beamform and finally the beam properties using Flocking. We use Lateral Inhibition and Flocking because they enable us to use local state information as opposed to other techniques. We support our work with simulation results and analysis, which show that a reduction of up to $40\%$ can be achieved for a high-density network. We also show the effect of $hopcount$ used to create regions on average path length, clustering coefficient and connectivity.
\end{abstract}


\section{Introduction}\label{sec:sec1}

Decades of academic and industrial research in wireless networks \cite{Akyildiz} has led to the tremendous growth of wireless networks requiring researchers to address manageability and scalability issues. Due to these issues, most of the research work has been oriented towards autonomous wireless networks. The autonomous behavior of the wireless nodes made decentralized computing and cost efficient topology deployment possible \cite{Dressler}. It was also proved that self-organization of the network can lead to better performance.

An attractive model to achieve better network performance is the Small World network. Small world networks are characterized by reduced Average Path Length ($APL$) and high Clustering Coefficient ($CC$). Here, the $APL$ is the mean of $hopcount$ between all pairs of nodes in the network. Consider a node, $v$, with $k$ neighbors. In the sub-graph of these $k+1$ nodes, the $CC$ is defined as the fraction of links that exist to the maximum number of links that could have existed in the sub-graph. Drawing inspiration from the experimental work of Stanley Milgram \cite{Milgram}, Watts et al \cite{Watts} proposed a model that could achieve small world properties. In the model, Watts et al proposed, small world properties could be reached by randomly rewiring a few existing links within the network. Watts et al showed that the dynamics of these small world networks lie between that of a regular network and a random network \cite{Watts,Wattsbook}. To prove the findings, however, Watts et al used a regular wired network and called the rewired links as shortcuts. Many complex real world networks such as internet, biological networks, food web and social networks also demonstrate small world properties \cite{Barabasi,BarabasiAlbert,Newman}. In real world networks where there is a non-uniform distribution of nodes, these real world networks were shown to exhibit the properties of scale-free networks marked by power law degree distribution. Section \ref{sub:subswn} provides more details on small world networks.

In a wireless ad hoc network, achieving small world properties can help us in many ways. Having a low $APL$ would increase the performance of the network in terms of communication \cite{Brust,KleinbergSWP} (reduced traffic per unit area, reduced congestion and reduced signal interference), low latency and reduce the overall energy consumption in the network during the data communication. On the other hand, maintaining the $CC$ would ensure connectivity to the neighborhood and would make the network resilient \cite{Albert,GuidoniLoureiro}. However, Watts' model cannot be applied directly to wireless ad hoc networks because of the spatial nature of such networks. In wireless ad hoc networks, addition of a shortcut between any two nodes should depend on the distance between two nodes. Helmy in \cite{Helmy} first studied the effect of adding few distance-limited links in the network. He showed that, upon introduction of distance-limited links, wireless ad hoc networks show small world properties. He concluded that, when the shortcut lengths are $\frac{1}{4}$th of the network diameter, there is a maximum reduction in the $APL$. Thus, proving that realization of small world properties in a wireless ad hoc network depends crucially on the length of shortcuts created among nodes. Another important factor in the realization of small world properties is the choice of nodes among which shortcuts are to be created. One method to obtain these nodes is that of \emph{preferential attachment} \cite{Simon,BarabasiAlbert}, typically observed in real world networks, wherein links are created to nodes with high structural importance. It was shown that, analogous to real world networks, using  \emph{preferential attachment} for creation of distance-limited links in a spatial network resulted in reduced network diameter \cite{BarthelemyMark,Manna}. This was accompanied by high clustering coefficient and a shift in the node degree distribution towards power law. These results motivate us to say that, creation of links to nodes having high structural importance in the network can result in the desired small world characteristics.


The creation of a wireless ad hoc network with the small world properties also depends on the manner in which distance-limited links are added. Such links can be added through different techniques like: 1) creating the directional beam using the same power as when the node was operating in the omnidirectional mode; 2) increasing the omnidirectional transmission range of the node; 3) introducing of few long wired links \cite{Sharma}; 4) introducing special nodes with higher omnidirectional transmission range deterministically in the network \cite{GuidoniLoureiro}; 5) using another antenna for beamforming in addition to the omnidirectional antenna.

\begin{figure}
    \centering
    \includegraphics[width=0.5\textwidth]{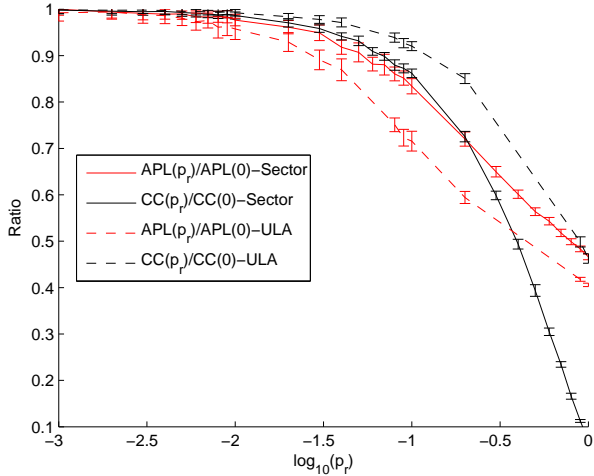}
    \caption{Source: \cite{Banerjee}, Effect of beamforming on the $APL$ and the $CC$ when the nodes are using 1) Sector model and 2) $ULA$ Model. The results obtained show a reduction in the $APL$ while almost no change in the $CC$ for the case when we use a realistic antenna model. On the other hand, for the theoretical model, the reduction in the $APL$ is relatively less while the reduction in $CC$ is considerably large. The number of nodes that beamform is shown with a probability value in the log scale. The results also show that the reduction in the $APL$ increases with an increase in the number of beamforming nodes, \cite{Watts,Wattsbook}. Here, $APL(p_r)$ and $CC(p_r)$ are the $APL$ and the $CC$ of the network when $p_r\%$ of the nodes create long-range links. $p_r=0$ means no node is beamforming. Further, in the figure, we normalize $APL(p_r)$ and $CC(p_r)$ to account for the variation in the $APL$ and the $CC$.}
    \label{fig:figinitial}
\end{figure}

Talking about the self-organization characteristics of the nodes, only techniques one and two mentioned above qualify. However, even though other techniques help in achieving desired network characteristics, they lack self-organization capabilities. In addition, the second technique suffers from the problem of early death of the node due to increased energy consumption. Thus leaving us only the first technique. Achieving reorganization or rewiring in a wireless ad hoc network through the first technique is hard due to the spatial nature of the wireless ad hoc network. Finding the beam direction, the beam length and determining the new neighborhood are primary issues associated with rewiring in a wireless ad hoc network. Our previous study, \cite{Banerjee}, proved that the use of distance-limited long links in wireless ad hoc network to achieve small world properties is beneficial, (Cf. Fig. \ref{fig:figinitial}).

Motivated by this, in this study, we investigate how we can increase connectivity, reduce the $APL$ and almost maintain the $CC$ in a non-uniformly distributed wireless ad hoc network. We thus propose an algorithm that achieves these goals by creating long-range directional beams between nodes that have low and high structural importance. The decentralized computing and self-organizing requirements of such an approach motivate us to draw inspirations from nature. We further propose that Lateral Inhibition \cite{Lawrence,Nagpal,NagpalMamei,Afek} and Flocking \cite{Reynolds}, in conjunction with the centrality concept of graph theory, can provide valuable insights in building a solution to our problem.

We use Lateral Inhibition to create small logical regions within a network. The use of Lateral Inhibition not only reduces the message complexity but also enables us to apply the Flocking rule analogy successfully. We use analogy of Flocking rules to identify the nodes that beamform and the beam properties. According to the rules, explained later in section \ref{sub:subflocking}, it is important to identify stray nodes, align the nodes and move them towards the centroid of their neighborhood. Analogous to this, after region formation in a non-uniformly distributed wireless ad hoc network, we use Flocking rules to identify the beamforming nodes and direct the beams of these beamforming nodes towards the centroid of the region. The centroid node in the region has a high structural importance. Beamforming towards the centroid node of the region contributes towards reducing the $APL$ because the centroid node of the region is the most connected node and has the highest Closeness Centrality measure. Thus, beamforming towards the centroid node is the \emph{preferential attachment} behavior of the beamforming node, thereby making centroid finding a prerequisite to Flocking. In a distributed system where nodes only have local information and lack $GPS$ facilities, exact centroid node identification of the region is challenging. We can only make an estimate to the centroid node location in the region. We, therefore, use the self-organizing virtual coordinate scheme combined with the centrality concepts to identify the centroid nodes.

Thus, our algorithm design is such that it first identifies regions using Lateral Inhibition, then identifies the centroid nodes of the regions and then uses the analogy of flocking rules to identify the nodes that will beamform along with their beam properties. Section \ref{sec:sec3} gives a formal description of our proposed algorithm.

The organization of rest of this paper is as follows. Section \ref{sec:sec3} presents the assumptions used for the proposed algorithm along with the algorithm specifications. Section \ref{sec:sec5} presents the formal definitions. Section \ref{sec:sec6} and \ref{sec:sec7} discuss the simulation setup and the results respectively. For the readers who are unfamiliar with the concepts used in this paper, we provide a detailed description of the same in section \ref{sec:sec2}. We finally conclude our work in the section \ref{sec:sec8} after providing insights to some future research directions in section \ref{sec:sec9}.

\section{Assumptions and Algorithm}\label{sec:sec3}

\subsection{Assumptions}
To address issues mentioned in the Introduction, we focus ourselves towards the deployment of homogenous and autonomous wireless ad hoc nodes with no central entity controlling the nodes. This type of deployment enables us to easily apply self-organizing features, achieve global consensus with very limited local information, make any eligible node the group leader, make the system highly fault tolerant, ease the topological maintenance, lower the deployment cost and extend to incorporate the mobility of the nodes in the future. Further, the nodes are set to have an omnidirectional transmission range $r$. We assume a non-uniform distribution of nodes generated using thinning process defined by Bettstetter et al \cite{BettstetterGyarmati}. The non-uniform distribution of nodes allows us to realize scenarios that are more realistic. The algorithm proposed by Bettstetter et al proceeds by removing nodes which have less than $\ell_{min}$ neighbors within a transmission range $r_{b}$ (ref. section \ref{sub:subnonunifdistrib}). Further, we assume the deployment of the nodes on a 2-D plane of area $A$.

As part of our network setup, our algorithm assumes each node to have an antenna consisting of $M$ isotropic elements. The use of single antenna element results into omnidirectional beam while use of more than one antenna element results into a long-range directional beam. A node, however, decides to use more than one antenna element using simple local rules mentioned later in this section. The nodes use beamforming only to transmit data but use omnidirectional beams for reception. We have used the Sector model\footnote[1]{Sector model approximates realistic antenna models} \cite{Yu} to visualize our algorithm and have assumed transmission of data to be synchronous.


Further, we assume that the nodes lack $GPS$ facilities and global network knowledge. To achieve our goal, it is thus first essential to know what information can be used by the nodes. We limit a node to use local information along with that of its one hop neighborhood. Determining single hop neighborhood to build the local information is thus essential for the correct operation of the algorithm. Various studies have proposed many neighborhood discovery mechanisms, eg. \cite{Vasudevan}, and have carefully analyzed them. Therefore, for our approach, we assume that all the nodes have information about their neighborhood.

\begin{table*}
\centering
    \begin{tabular}{|l|l|l|l|}
        \hline
        \textbf{Notation} & \textbf{Meaning} & \textbf{Notation} & \textbf{Meaning}\\
        \hline
        $A$ & simulation area & $g$ & gradient\\
        $G$ & network with set of vertices $V$ and set of & $g_{max}$ & maximum gradient\\
            & edges $E$ & $e\_bet_{v}$ & Egocentric Betweenness of $v$ w.r.t. its\\
        $G_{i}$ & region $G_{i}|G_{i} \subset G$ with set of & & cluster\\
                & vertices $V_{i}$ and set of edges $E_{i}$ & $hops(v,w)$& $hopcount$ between node $v$ and $w$\\
        $N$ & number of regions formed & $v_{i}(x, y)$ & virtual coordinates of $v$ in the region $G_{i}$\\
        $v$ & node $|v \in V$ & $v_{i}(x^{*}, y^{*})$ & updated virtual coordinates of $v$ in the\\
        $v_{i}$ & node $v$ in region $G_{i}|v_{i} \in V_{i}$ & & region $G_{i}$\\
        $r$ & transmission radius & $\varepsilon$ & error margin\\
        $r_b$ & Bettstetter transmission radius & $M$ & max antenna elements available with $v$\\
        $\rho$ & average node density & $m$ & number of antenna elements used by $v$\\
        $ID_{v}$ & identification number of node $v$& & to beamform $| m\in [2,M]$\\
        $L_{v}$ & neighbor list of $v|v\in V$ & $RC_{v}$ & set of centroid nodes reachable from $v$\\
        $L_{v,i}$ & neighbor list of $v$ in the region $G_{i}|v\in V_{i}$ & & with their $hopcount$ that are within $g_{max}$\\
        $\ell_{min}$ & minimum number of neighbors used for & & hops from $v$ when $v$ is not beamforming\\
                  & creating a non-uniform distribution & $RC_{v}^{*}$ & set of centroid nodes reachable from $v$\\
        $deg_{v}$ & size of $L_{v}$, i.e., degree of $v$ & & with their $hopcount$ when $v$ is beamforming\\
        $H$ & set of all region heads & $\theta$ & beam direction, i.e., the sector\\
        $h_{i}$ & head node of the region $G_{i}|h_{i} \in H$ & $B_{b}$ & boresight direction\\
        $C$ & set of all centroid nodes & $B_{l}$ & beam length\\
        $c_{i}$ & centroid node of the region $G_{i}|c_{i}\in C$ & $B_{w}$ & beam width\\
        $P$ & set of all peripheral nodes & $APL$ & Average Path Length\\
        $P_{i}$ & set of peripheral nodes in the region $G_{i}|P_{i}\in P$ & $CC$ & Clustering Coefficient\\
        $\wp_{i}$ & peripheral node $|\wp_{i}\in P_{i}$&$ULA$ & Uniform Linear Antenna Array\\
        $\varrho_{\wp_{i}}$ & peripheral neighbor of $\wp_{i}|\varrho_{\wp_{i}}\in P_{i}$&$GSCC$ & Giant Strongly Connected Component\\
        $S$ & Set of nodes neither in $C$ nor in $P$&$GIN$ & Giant In Component\\
%
%
        \hline
    \end{tabular}
    \caption{Notations and their meaning.}
    \label{table:table3}
\end{table*}

It is also essential to address the self-organizing paradigms, \cite{Prehofer}, to claim for the self-organizing behavior of the network. Prehofer et al's \cite{Prehofer} paradigms state: designing local rules to achieve global properties, implicit coordination, minimizing the use of historic information about the state of the network and designing an algorithm that changes with environment parameters. Our algorithm uses only locally available information to determine the beamforming nodes, beam properties and the regions. The nodes implicitly coordinate with their neighbors to determine the node with the highest $hopcount$ from the centroid of the region. For a given region, the nodes also coordinate implicitly to determine the centroid node of that region. The current discussion focuses on a static network. In dynamic network scenarios, optimizing the extent of reconfiguration to deal with frequent changes in state information is likely to be a crucial factor. We leave this for future investigation but offer some insights in section \ref{sec:sec9}.

We further describe the system model and the algorithm in the following sections.

\subsection{System Model}

Given a network, $G(V,E)$, where $V$ is the set of vertices and $E$ is the set of edges, we visualize $G$ as a network consisting of $N$ logical regions, $\{G_{1},G_{2},\dots,G_{N}\}$, i.e., $G=\bigcup_{i=1}^{N}{G_{i}}$. Each region, $G_{i}$, consists of the set of nodes, $V_{i}|V_{i}\subset V$ and $V=\bigcup_{i=1}^{N}{V_{i}}$, and set of edges, $E_{i}|E_{i}\subset E$ and $E=\bigcup_{i=1}^{N}{E_{i}}$. All vertices in $G_{i}$ are located within $g$ hops of a head node, $h_{i}$. As a part of our algorithm, we use Lateral Inhibition to identify regions and regional heads.

We characterize the set of vertices, $V$, into three sets. These are termed as the Peripheral node set, the Centroid node set and the Standard node set. We provide separate role to the nodes in these sets. The Peripheral nodes set ($P$) contains the nodes that beamform. The Centroid node set ($C$) contains the nodes towards which the nodes in the Peripheral node set beamform. We call the set of remaining nodes, $S=V-(P\bigcup C)$, as the Standard node set. Further, we call nodes in these sets as the peripheral nodes, the centroid nodes and the standard nodes respectively.

Mathematically, Closeness centrality of a node, $v \in V$, in a graph $G$ is equal to $\frac{1}{\sum_{w \in V}{hops(v,w)}}$, where $hops(v,w)$ is the $hopcount$ between nodes $v$ and $w$. The node having maximum Closeness Centrality is the centroid of the graph and has a high structural importance. For the vertex sets defined above, nodes in the set $P$ have lowest value of Closeness Centrality, i.e., $\operatorname*{arg\,max}\limits_{v\in V}\{\sum_{w \in V,v\neq w} hops(v,w)\}$. However, the nodes in the set $C$ have highest value of closeness centrality, i.e., $\operatorname*{arg\,min}\limits_{v\in V}\{\sum_{w \in V,v\neq w} hops(v,w)$\}. A node in $P$ beamforms towards a node in $C$ in order to minimize the distance to other nodes and reduce $APL$.

The directional beam is modeled using Sector model, i.e., for a given directional beam length $B_l$, the corresponding beam width, $B_w$, is

\begin{equation}\label{eq:blbw}
    B_{w}=\frac{2\pi r^2}{B_{l}^{2}}
\end{equation}

In realistic antenna model, as beam length of the directional antenna is dependent on the number of antenna elements used, $m$, the corresponding value of $B_{l}$ used is $B_{l}=m*r$.

Further, table \ref{table:table3} lists the notations used in this paper.

\subsection{Algorithm}

We divide our approach into two parts:

\begin{itemize}
    \item[A)] Use of Lateral Inhibition technique and self-organizing virtual coordinate scheme for the identification of regions and the centroid nodes of the regions, so that there are less message overheads and nodes can beamform towards the centroid node to achieve reduced $APL$. Section \ref{sub:s1} provides more details.
    \item[B)] Use of flocking rules to identify the nodes that beamform, to determine beam properties that realize small world properties and improve connectivity. Section \ref{sub:s2} provides more details.
\end{itemize}

We describe these parts in detail in the next sub sections.

\subsubsection{Region formation and Centroid finding}\label{sub:s1}

The Closeness Centrality \cite{Freeman,Freemanlc} identifies the structural importance of the node in the network. The node with the highest Closeness Centrality value is the most central node in the network. Through this node, the spread of the information to other nodes is quick. To determine the Closeness Centrality of the node, the node requires the knowledge of other nodes in the region as suggested by the definition of Closeness Centrality, (ref. section \ref{subsubsec:closenessc}). This makes the Closeness Centrality a global measure. Storing information about all the nodes in the network can consume a lot of node's memory. When there is lack of global information, gathering such information can also be time consuming and the message complexity could be high. To overcome these problems, we create small logical regions. The creation of regions not only reduces the message complexity of the network but also reduces the effect on the $APL$ due to the failure of a node, thereby making the network more manageable, efficient and tolerant to failures \cite{BrustRibeiro}. Some algorithms designed in this direction were centralized. The Base Station chose the region heads based on the energy and the position of the nodes. Other techniques use either the transmission power or the degree or the mobility, eg., $WACA$ \cite{BrustAndronache}. On the contrary to centralized approaches, some algorithms were either distributed, \cite{Heinzelman}, or probabilistic \cite{Younis}.

We thus divide this part into two, identification of regions using Lateral Inhibition and identification of centroid node in the region. As we only have local information, we use degree of the node in the Lateral Inhibition process.

\begin{figure}[ht]
    \centering
    \includegraphics[width = 0.33\textwidth]{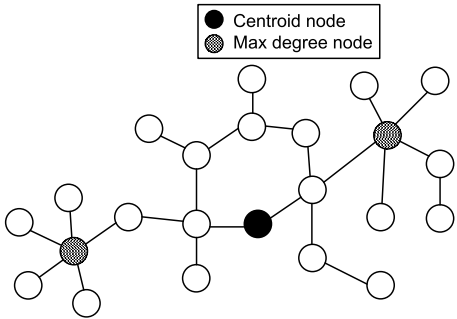}
    \caption{The max degree nodes are not at the center of the region. The Closeness Centrality of these nodes is less.}
    \label{fig:f1}
\end{figure}

\begin{figure*}
    \centering
    \mbox
    {
        \subfigure[Uniform Node Distribution.]
        {
            \includegraphics[width = 0.38\textwidth]{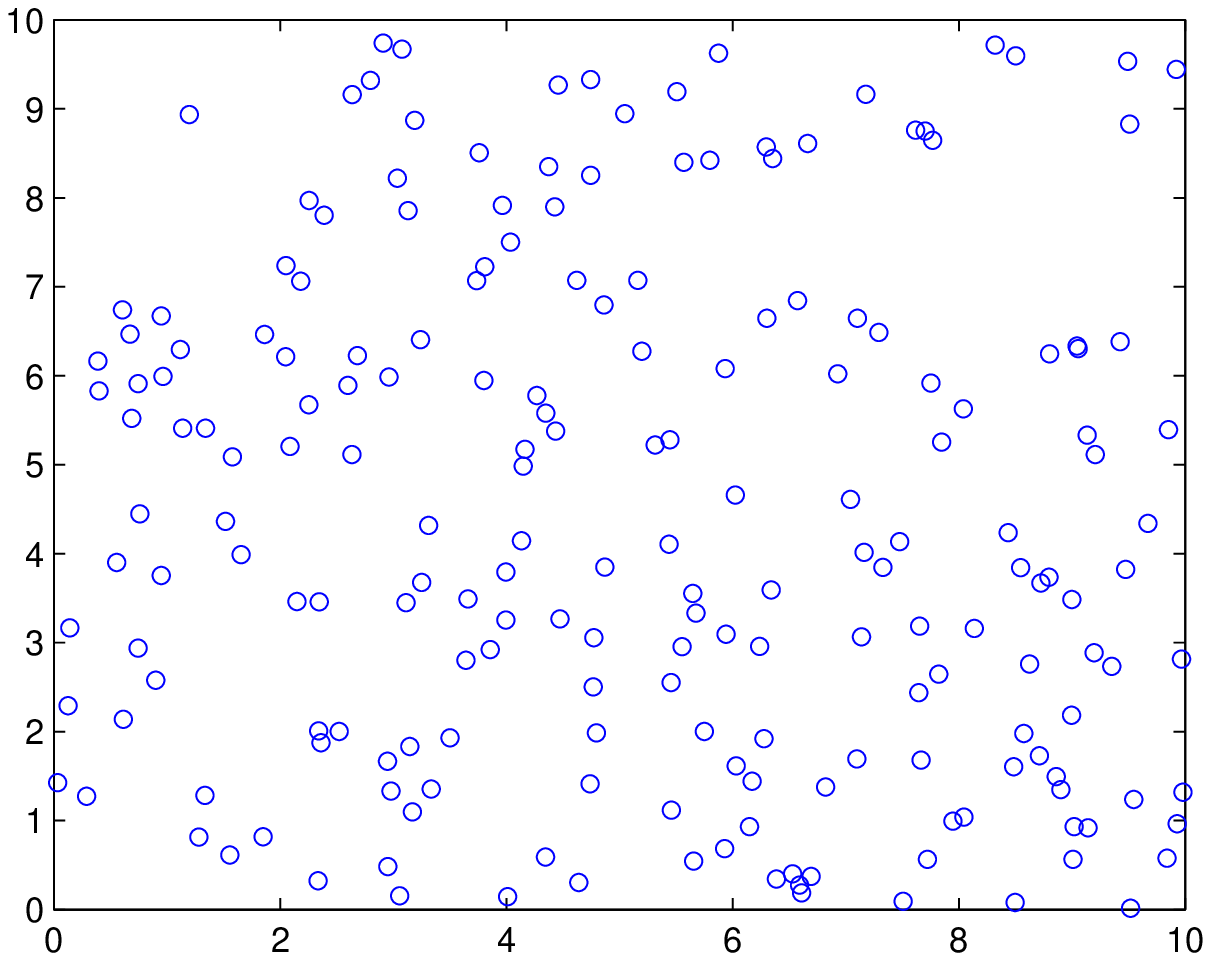}
            \label{subfig:fignodedistbunif}
        }
        \hspace{0.5cm}
        \subfigure[Distribution after applying Thinning process with $r_b=1$ and $\ell_{min}=5$.]
        {
            \includegraphics[width = 0.38\textwidth]{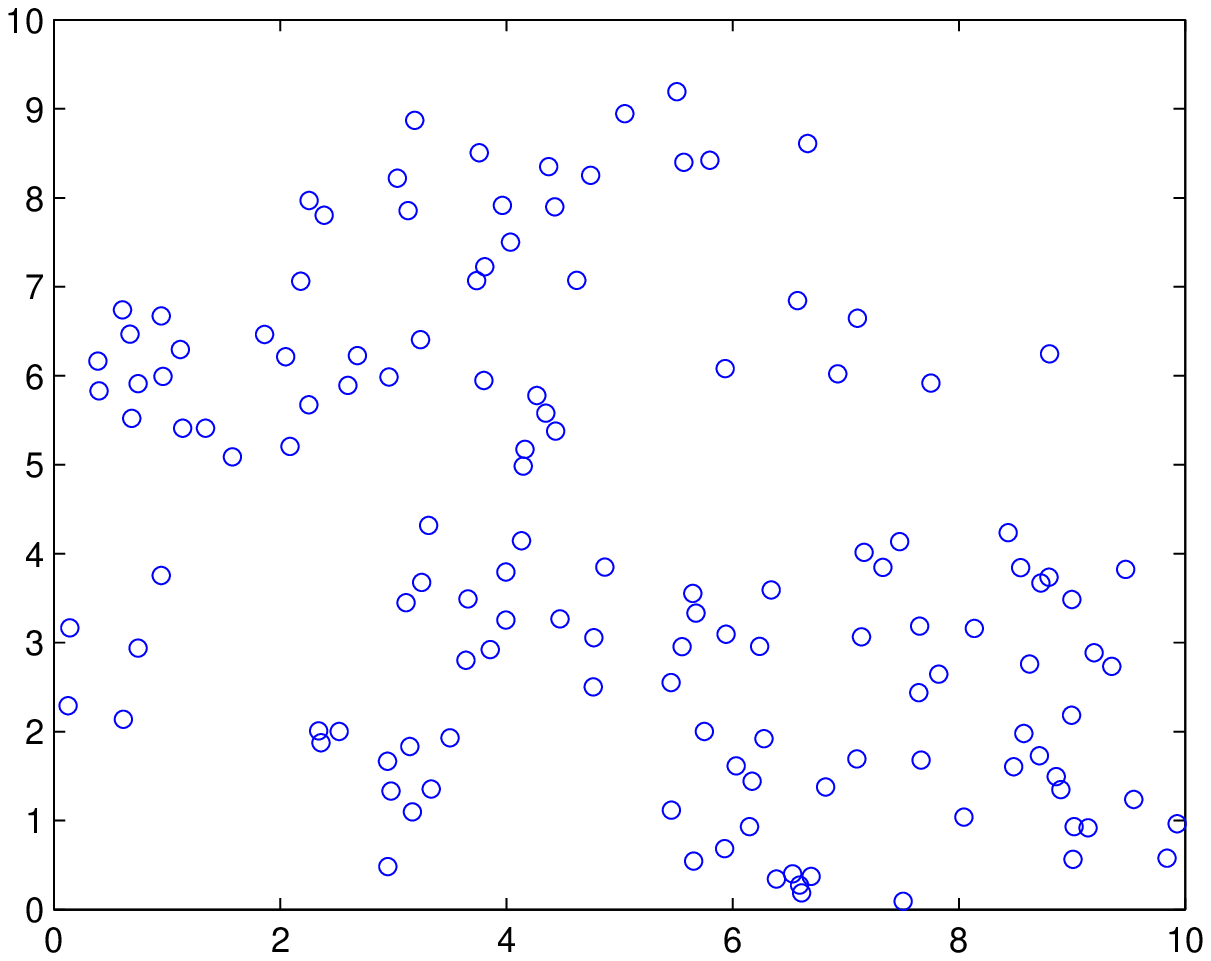}
            \label{subfig:fignodedistbnonunif}
        }
    }
    \mbox
    {
        \subfigure[Identified regions in the deployment shown by the Fig. \ref{subfig:fignodedistbnonunif}.]
        {
            \includegraphics[width = 0.38\textwidth]{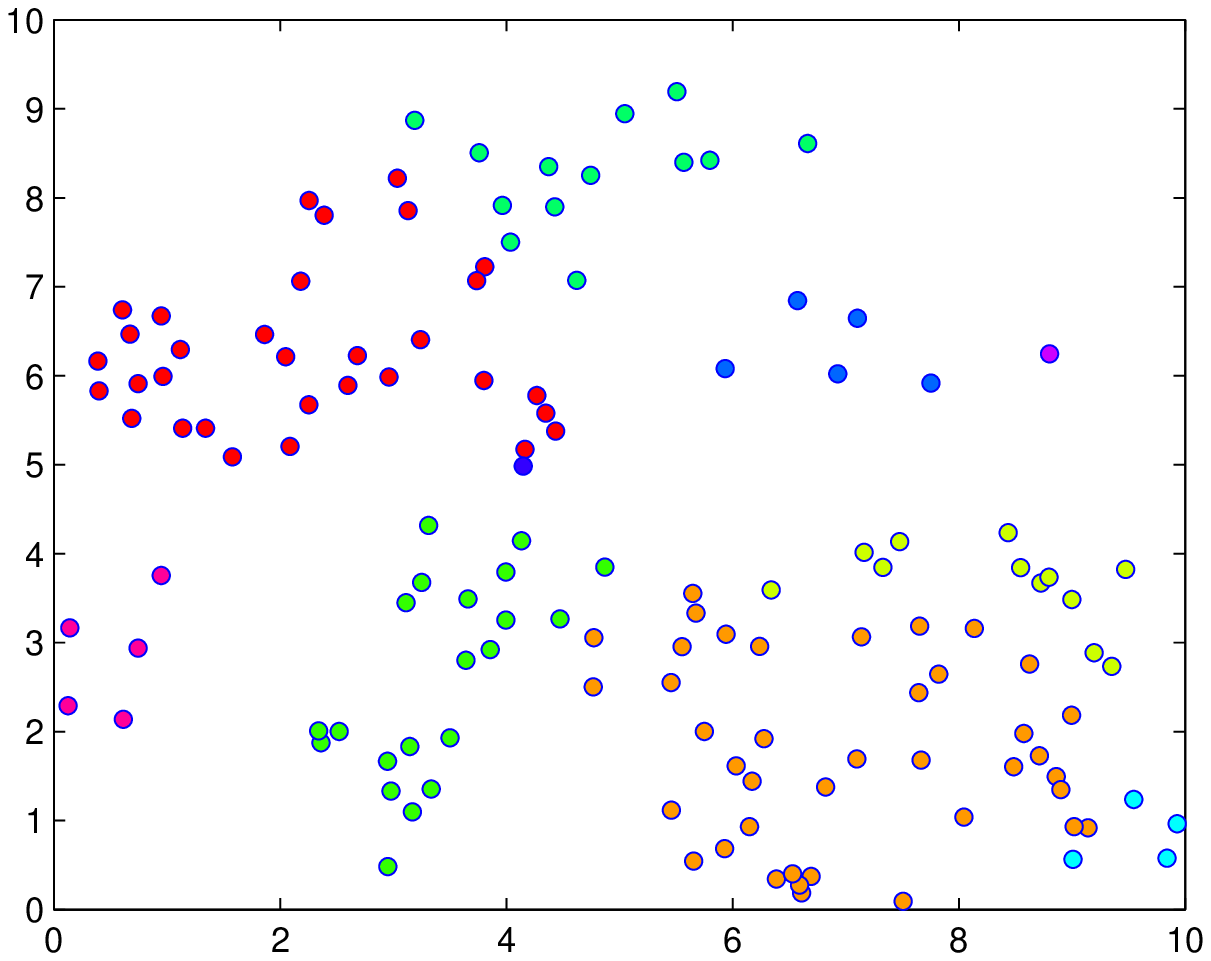}
            \label{subfig:regions}
        }
        \hspace{0.5cm}
        \subfigure[Identified uninhibited nodes created using the \cite{Afek} algorithm for the deployment shown by the Fig. \ref{subfig:fignodedistbnonunif}. As there is only one head in the region, the number of uninhibited nodes directly refers to the number of regions created. The nodes shown with + are the uninhibited nodes while the nodes shown with o are the inhibited nodes.]
        {
            \includegraphics[width = 0.38\textwidth]{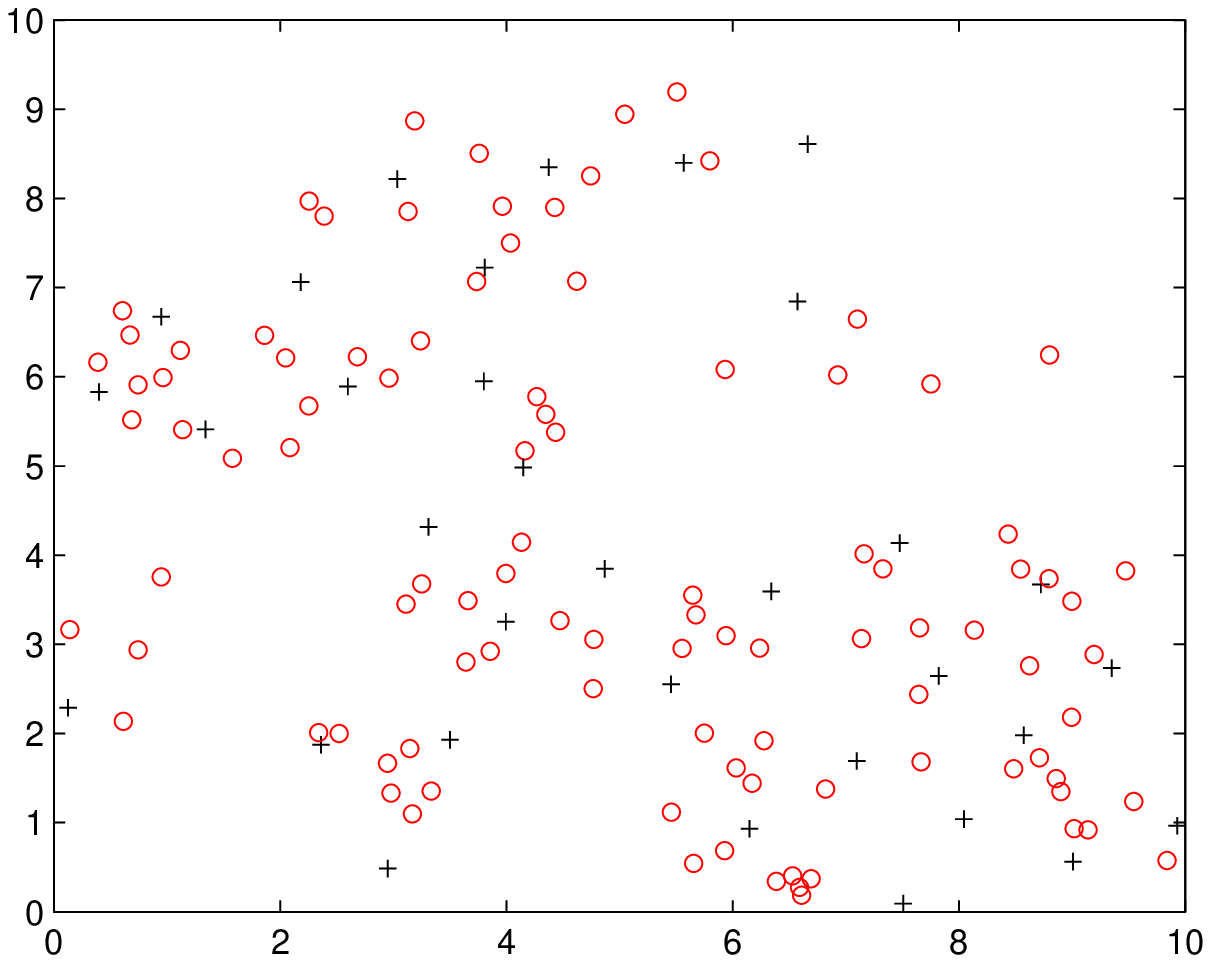}
            \label{subfig:regionsafek}
        }
    }
    \mbox
    {
        \subfigure[The gradient of the nodes created using the $hopcount$ for the regions created in the Fig. \ref{subfig:fignodedistbnonunif}. The peaks show the centroid nodes while the valley show nodes with the max gradient value.]
        {
            \includegraphics[width = 0.38\textwidth]{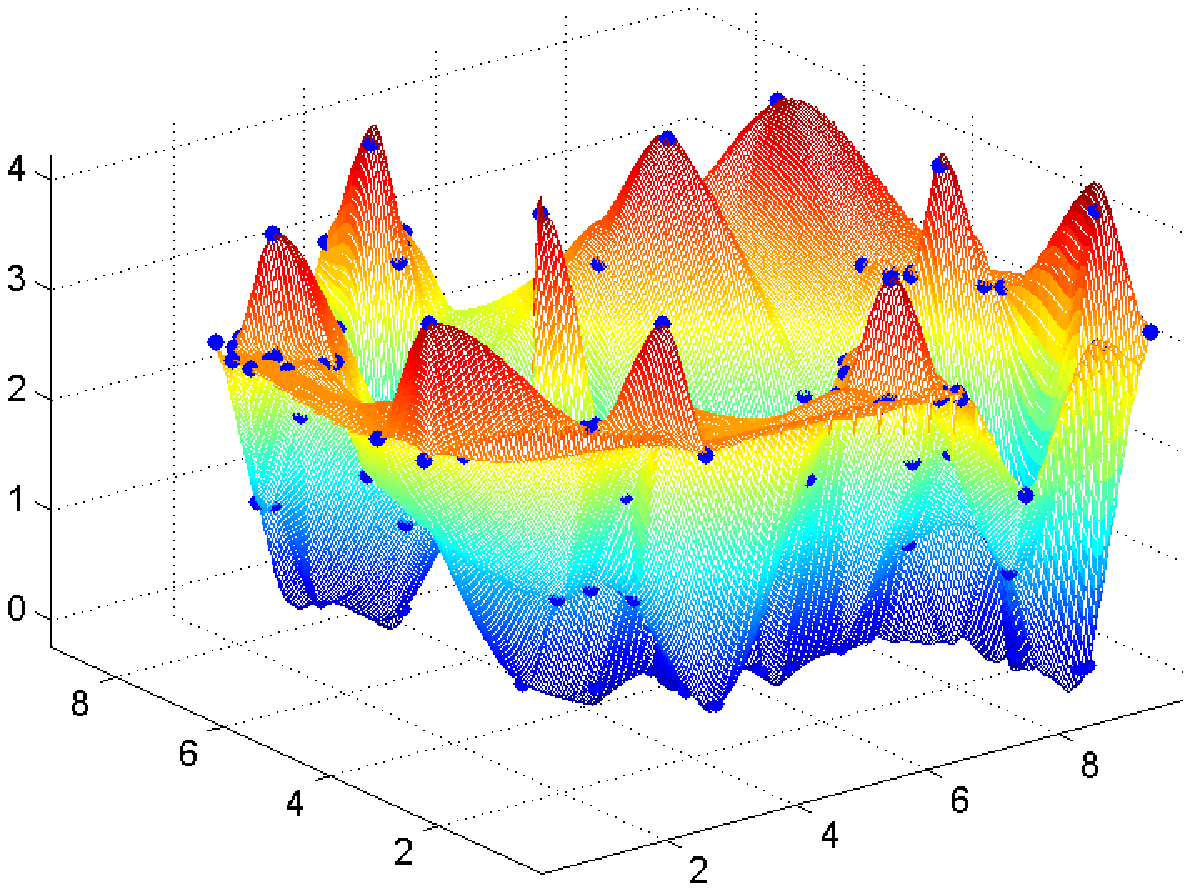}
            \label{subfig:gradient}
        }
        \hspace{0.5cm}
        \subfigure[Association of the nodes to the centroid nodes. The centroid of the region is marked with a black square.]
        {
            \includegraphics[width = 0.38\textwidth]{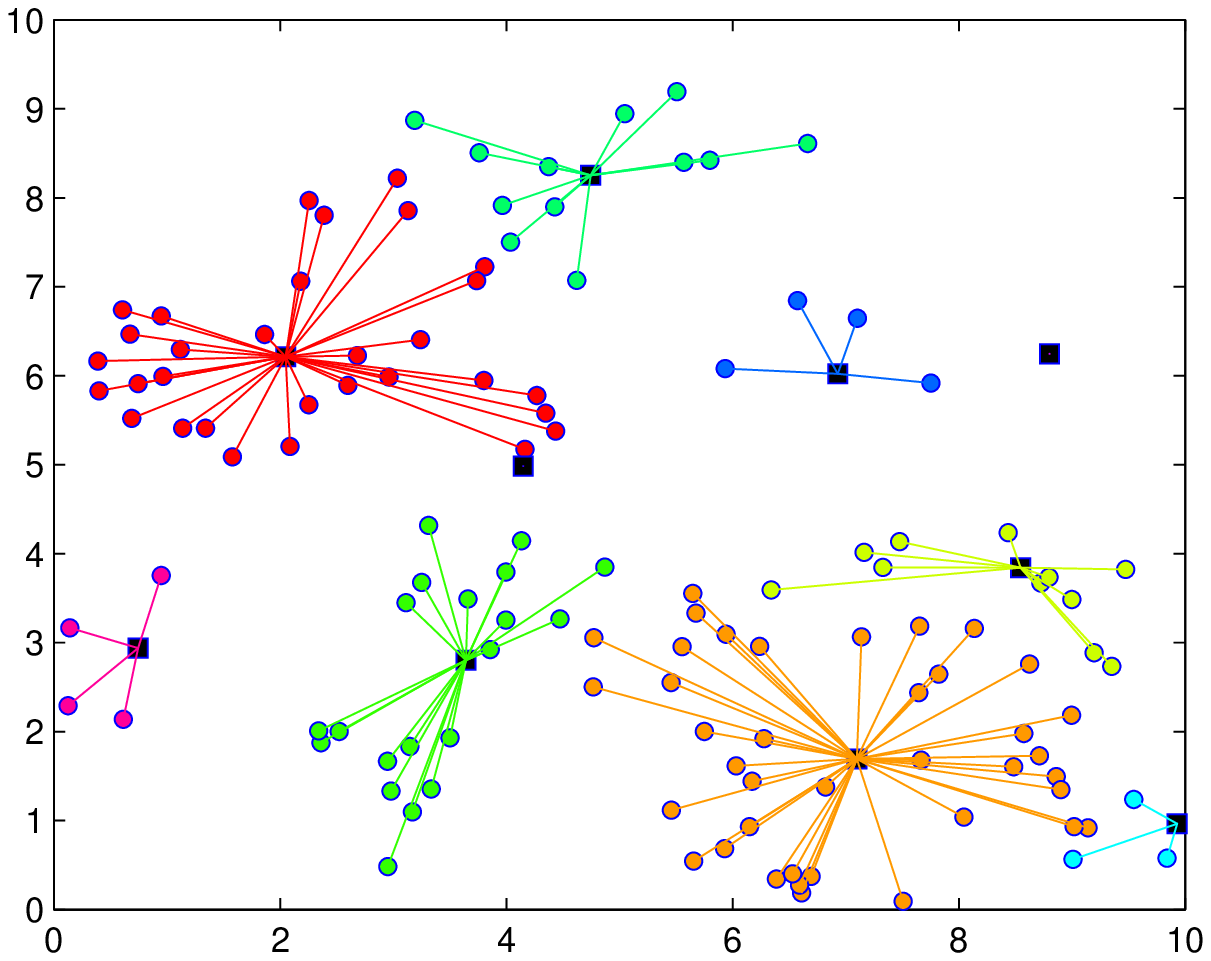}
            \label{subfig:centroidnodes}
        }
    }
    \caption{Region Formation and Centroid identification using $g<5$.}
    \label{fig:RegionFormation}
\end{figure*}

For Lateral Inhibition, we consider that a node $v_{i}$ broadcasts and stores a message containing following information: the identity of the head node to which $v_{i}$ is associated ($h_{i}$), its $hopcount$ from $h_{i}$ and the degree of $h_{i}$ ($deg_{h_{i}}$), where $v_{i}\in V_{i}$. Initially, all the nodes, $v\in V$, consider themselves as heads, i.e. $H=V$, and store their own information, i.e., $h_{i}=v$, $hopcount=0$ and $deg_{h_{i}}=deg_{v}$. Each node, $v\in V$, then broadcasts this information to its neighbors, $L_{v}$. Similarly, $v$ receives information from each of its neighbors and subsequently updates the information stored in it. Thus, a node replaces its stored values, if the stored degree, $deg_{h_{i}}$, is less than that of the received value and $hopcount+1$ is less than $g$, where $g$ is the gradient or the desired size of the regions. Further, if the stored and the received $deg_{h_{i}}$ are same, the node decides to update the stored information based on lower $hopcount$ value. If the $hopcount$ is also same, then the node randomly decides to update the stored information to received information. The node $v$ then broadcasts the updated information after incrementing the $hopcount$ by 1. Subsequently, $v$ removes itself from $H$, i.e., $H=H-\{v\}$, and inhibits itself from acting as the regional head. The process continues until all the nodes within $g$ hops from the maximum degree node reach a consensus about the head node. Due to $g$, the algorithm assigns same $h_{i}$ to all the nodes within $g$ hops of the head node. We call the nodes having same $h_{i}$ to belong to one region, $G_{i}$. The nodes lying at different $hopcount$ from the $h_{i}$ virtually creates a gradient of different hops around $h_{i}$, (Cf. Fig. \ref{subfig:gradient}). In the end, the algorithm tags a node with no neighborhood as the head as it has remained uninhibited, (Cf. Fig. \ref{subfig:regions}). The regions created differ from other Lateral Inhibition algorithms, \cite{Afek}, in a way that our algorithm creates regions that are not limited to 1 hop, (Cf. Fig. \ref{subfig:regions} and Fig. \ref{subfig:regionsafek}). However, the Lateral Inhibition technique does not guarantee that the head nodes identified above have a high Closeness Centrality value and are the most central nodes, (Cf. Fig. \ref{fig:f1}).

We thus now describe the steps for the centroid node identification in a given region, $G_{i}$, created using Lateral Inhibition described earlier. Due to the global properties of the Closeness Centrality and unavailability of any $GPS$ facilities within the nodes, we take insights from existing algorithms on self-organizing virtual coordinate systems. In self-organizing virtual coordinate system, the nodes identify their own coordinates relative to their neighborhood in the network. We however, make use of self-organizing virtual coordinate system to calculate centroid of the region. Existing techniques on self-organizing virtual coordinate system include \cite{Capkun,Caruso,Leong,Dabek,Rao,Priyantha,Watteyne,Awad}. These studies deploy various mechanisms to reach consensus. We use a method for achieving consensus on centroid location based on self-organizing virtual coordinate techniques that rely on averaging of local neighborhood values \cite{Rao,Watteyne}. This allows us to limit the information required to a single hop, and thereby have minimum communication overheads.

Thus, in our algorithm, all nodes $v_{i}\in V_{i}$ in $G_{i}$ assign themselves randomly selected virtual $xy$ coordinates, $v_{i}(x, y)$. The identity of the nodes in the virtual coordinate system, however, remains the same. The nodes then communicate to their neighbors in $G_{i}$ these coordinates, i.e., $L_{v,i}$. Using the coordinates of their local neighborhood, the nodes compute an average of the coordinates, $v_{i}(x^{*}, y^{*})$, and broadcast the average coordinates to their neighbors. The neighbors in turn use these coordinates to compute a new average. This process continues until all nodes in the region reach consensus of having same average $xy$ coordinates of the centroid.


The self-organizing virtual coordinate technique reveals the location of the centroid node in the self-organizing virtual coordinate system but not the identity of the node that is to be termed as centroid. In order to identify the centroid node of the region, nodes use their initially assigned virtual coordinates and the newly found average $xy$ coordinates. Each node $v_{i}$ checks if $v_{i}(x, y)=v_{i}(x^{*}, y^{*})\pm\varepsilon$, where $\varepsilon$ is the error margin, and declares itself as the centroid. This process might result into multiple nodes declaring themselves as the centroid as two or more nodes can lie within the $\varepsilon$ range of $v_{i}(x^{*}, y^{*})$. To avoid this, a node also considers its Degree and Egocentric Betweenness\footnote[2]{Egocentric Betweenness approximates the Socio-Centric Betweenness very well in the absence of global knowledge \cite{Marsden}}. The nodes within $\varepsilon$ range of $v_{i}(x^{*}, y^{*})$ share this information among themselves. Subsequently, the node having maximum sum of Degree and Egocentric Betweenness declares itself as the centroid of the region. As the node has same identity in the self-organizing virtual coordinate system as in the real coordinate system, the centroid node in the self-organizing virtual coordinate system will also be the centroid in the real coordinate system. After the identification of the centroid nodes, the centroid nodes broadcast their information in the network. All nodes then update their stored head information to their respective $c_{i}$'s and the $hopcount$ to $hops(v_{i},c{i})$.

This broadcasting of the centroid node information enables the nodes to build $RC_{v}$ for future use. $RC_{v}$ is the set of centroid nodes within $g_{max}$ hops of the node $v$, where $g_{max}>g$. Algorithm \ref{Algo1} represents the algorithmic description of the region formation and the centroid identification process. The Fig. \ref{subfig:centroidnodes} shows the centroid nodes for the regions identified in the Fig. \ref{subfig:regions}.

\subsubsection{Beamforming}\label{sub:s2}

In this part, we describe the steps involved in beamforming. According to the results of \cite{Helmy}, it requires only a small fraction of nodes with long link capabilities to achieve small world properties. In a self-organizing environment where all nodes possess beamforming capabilities, it is essential to identify nodes that create long-range beams along with the direction and the width of the beam. Flocking provides us with valuable insights in determining the answers to these questions. We use insights from the \textbf{Alignment} rule of Flocking to identify the set $P$. Alignment in Flocking is the change in the direction of the node to match its neighbors, in other words the change in the orientation of the node. Further, Alignment rule is, the node has to decide to change the direction and has to find the new direction. We modify the Alignment rule and say that our Alignment rule is only limited to the decision of whether to create the beam or not. The Alignment rule we apply is, thus, to identify the set of peripheral nodes, $P_{i}$ in the region $G_{i}$. Our algorithm uses the $hopcount$ of the neighborhood nodes to decide whether or not the node is a peripheral node, $\wp_{i}$, of the region $G_{i}$. If all $L_{v,i}$ of the node $v_{i}$ have $hopcount$ less than or equal to the node's $hopcount$ to the $c_{i}$, then the node declares itself as a peripheral node. i.e., for a given region $G_{i}$ with centroid $c_{i}$, $\wp_{i} \in P_{i} \iff hops(\wp_{i},c_{i})\geq hops(L_{\wp_{i}},c_{i})$. This implies that, a single unconnected node will become a peripheral node because it does not have any neighborhood. Further, we can also infer that two peripheral nodes can be neighbors of each other due to the equality in the condition.

\begin{algorithm}[!ht]
    \caption{Region formation and centroid finding}\label{Algo1}
    \begin{algorithmic}[1]
        \STATE Let \ensuremath{U=uninhibited};
        \STATE Let \ensuremath{I=inhibited};
        \STATE Let \ensuremath{ID=identity\; of\; node};
        \STATE $\backslash\backslash$ Region formation;
        \FORALL {$v \in V$}
            \STATE set \ensuremath{v_{Status}=U}
            \STATE set \ensuremath{v\_coordinates=v_{i}(x,y)}
            \STATE Initially broadcast(\ensuremath{ID_{v}, hopcount=0, deg_{v}})
        \ENDFOR
        \REPEAT
            \STATE \ensuremath{recv}=receive(\ensuremath{ID, hopcount+1, degree})
            \IF {\ensuremath{deg_{v}<degree\;\&\; hopcount<g}}
                \STATE \ensuremath{v_{Status}=I} \& broadcast(\ensuremath{recv})
            \ENDIF
        \UNTIL {converges}
        \STATE $\backslash\backslash$ Centroid finding;
        \FORALL {\ensuremath{v_{i}\in V_{i}\in V}}
            \STATE \ensuremath{v_{i}(x^{*},y^{*})}=Cent\_finding(\ensuremath{v_{i}(x,y), L_{v,i}(x,y)})
        \ENDFOR
        \FORALL {\ensuremath{v_{i}\in V_{i}\in V |v_{i}(x,y)-\varepsilon<v_{i}(x^{*},y^{*})<v_{i}(x,y)+\varepsilon}}
            \STATE compute \ensuremath{sum_{v_{i}}=sum(deg_{v_{i}}, e\_bet_{v_{i}})}
        \ENDFOR
        \FORALL {\ensuremath{v_{i}\in V_{i}\in V}}
            \STATE \ensuremath{c_{i}=v_{i}|v_{i}=max\{sum_{v_{i}}\}}
            \STATE \ensuremath{C=C+{v_{i}}}
        \ENDFOR
        \FORALL {\ensuremath{v \in V}}
                \STATE formulate \ensuremath{RC_{v}}
        \ENDFOR
    \end{algorithmic}
\end{algorithm}

\begin{figure}[ht]
    \centering
    \mbox
    {
        \subfigure[The difference in the beam properties when a $\wp_{i}$ uses different number of antenna elements.]
        {
            \includegraphics[width = 0.33\textwidth]{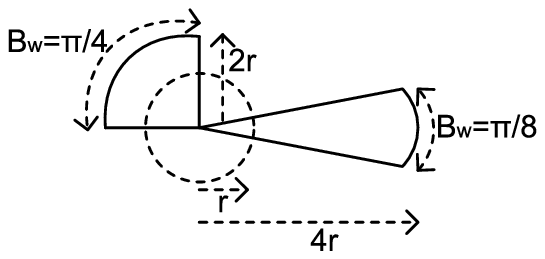}
            \label{subfig:fig11}
        }\quad
    }
    \mbox
    {
        \subfigure[Connectivity when 2\newline antenna elements are used.]
        {
            \includegraphics[width = 0.22\textwidth]{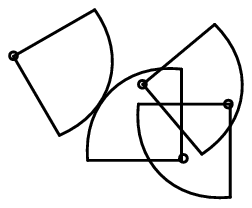}
            \label{subfig:fig12}
        }\quad
        \subfigure[Connectivity when 4\newline antenna elements are used.]
        {
            \includegraphics[width = 0.22\textwidth]{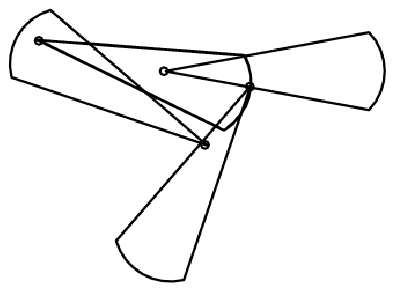}
            \label{subfig:fig13}
        }\quad
    }
    \caption{Relationship between beam properties and connectivity.}
    \label{fig:figsectormodel}
\end{figure}

The peripheral nodes randomly choose the number of antenna elements, $m \in [2,M]$, and use the above rules to beamform. Considering $B_{l}$ to be equal to $m*r$ in a Sector model, by keeping constant power as used for omnidirectional beam, we can easily compute $B_{w}$ from eq. (\ref{eq:blbw}) as $B_{w}=\frac{2\pi}{m^2}$. From this we infer that, to cover all the directions, minimum number of sectors that we need to consider is $m^2$. The dependency of $B_{l}$ and $B_{w}$ on $m$ affects the connectivity of the network. The Fig. \ref{fig:figsectormodel}(a) shows the variation in $B_{l}$ and $B_{w}$ when $m>1$. When $B_{l}$ is smaller, i.e., when we use less number of antenna elements, the probability of connecting to the neighbors is high as the beam is wider, (Cf. Fig. \ref{fig:figsectormodel}(b)). However, when $B_{l}$ is longer, i.e., when we use more antenna elements, the probability of connecting to a neighbor is low as the beam is narrower, (Cf. Fig. \ref{fig:figsectormodel}(c)).

As the number of sectors increase exponentially with an increase in the number of antenna elements, there is an increase in the time taken to decide the best sector. Checking all the sectors formed for all $m\in[2,M]$ requires a test of $\frac{(M)(M+1)(2M+1)}{6}-1$ sectors. The complexity of such a test is $O(M^3)$. This results into more energy consumption at the node. To reduce this energy consumption and the complexity to $O(M^2)$, our algorithm randomly selects the number of antenna elements, $m\in[2,M]$, and only tests the corresponding set of $m^{2}$ sectors.

\begin{figure}[ht]
    \centering
    \includegraphics[width = 0.45\textwidth]{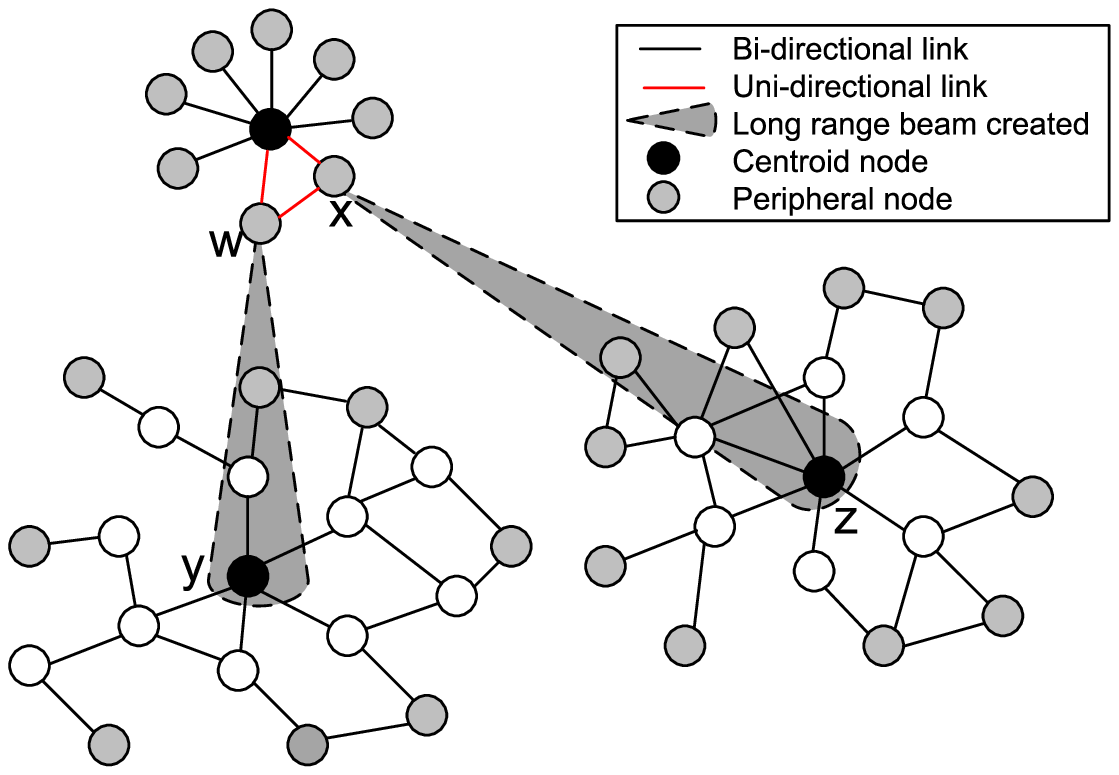}
    \caption{Nodes beamform in different directions. Two peripheral nodes $w$ and $x$ which were initially neighbors of each other, create beams in different directions. In order to have increased connectivity, the node $w$ creates a beam towards the region containing the centroid node $y$, while the node $x$ creates a beam towards the region containing the centroid node $z$. The maximum gradient value for Lateral Inhibition is 4.}
    \label{fig:f2}
\end{figure}

Non-uniformity reduces the size of the giant component in the wireless ad hoc network. It is thus important for the nodes to find different network components and connect them using beamforming. \textbf{Separation} rule of Flocking provides us insight towards this problem. Separation rule states that the nodes should maintain certain distance with their neighbors. Our algorithm applies similar analogy to address the connectivity issue. We say, in order to increase connectivity, nodes create beam in different directions from their peripheral neighbors. Consider $\varrho_{\wp_{i}} \in P_{i}$ as a peripheral neighbor of $\wp_{i}$ then for all $\varrho_{\wp_{i}}$'s, $\wp_{i}(B_{b}) \neq \varrho_{\wp_{i}}(B_{b})$ must hold. Here $B_{b}$ is the boresight direction. To make this decision, if $\varrho_{\wp_{i}}$ of a $\wp_{i}$ decides to create the beam in certain direction, $\varrho_{\wp_{i}}$ informs $\wp_{i}$ about the chosen direction before it actually creates the beam. $\wp_{i}$ then tries to create the beam in another direction. Further, $\wp_{i}$ gives preference to connect to the nodes in other region rather than that of its own. This increases the possibility of connecting to an isolated region. The Fig. \ref{fig:f2} shows two node $w$ and $x$ which were initially neighbors of each other, create beams in different direction in order to increase connectivity.

Nevertheless, we still have to address the best direction of the beam and the knowledge of whether a $\wp_{i}$ has a node within its 1 hop. We address these problems next in this section.

To the above-mentioned problem, we use analogy of \textbf{Cohesion} rule of Flocking to determine the best direction of the beam. In Flocking, Cohesion rule states that a node should move towards the centroid of the neighborhood to remain connected to all of its neighbors. We apply this definition of Cohesion in our algorithm because we want to bind a peripheral node with other nodes in minimum hops. From the previous section, we already know that the centroid node has the highest Closeness Centrality value in a given region. Directing the peripheral node's beam towards the centroid node would help reduce the average distance of the peripheral node to other nodes of the region in which the centroid node lies.

Combining Separation and Cohesion rules as discussed above, we can say that, if the centroid node chosen by the peripheral node and the peripheral node itself were not connected initially, connecting them would help in increasing the connectivity, (Cf. Fig. \ref{fig:f2}). On the other hand, if the centroid node chosen by the peripheral node was within some hops from the peripheral node, it will lead to the reduction in the $APL$.

\begin{figure*}[ht]
    \centering
    \mbox
    {
        \hspace{0.4cm}
        \subfigure[One component with three regions when $g=3$. Here, the node $x$ can create the beam towards $y$ or $z$, but because the $hopcount$ to $y$ is more than $z$, node $x$ creates the beam towards $y$.]
        {
            \includegraphics[width = 0.31\textwidth]{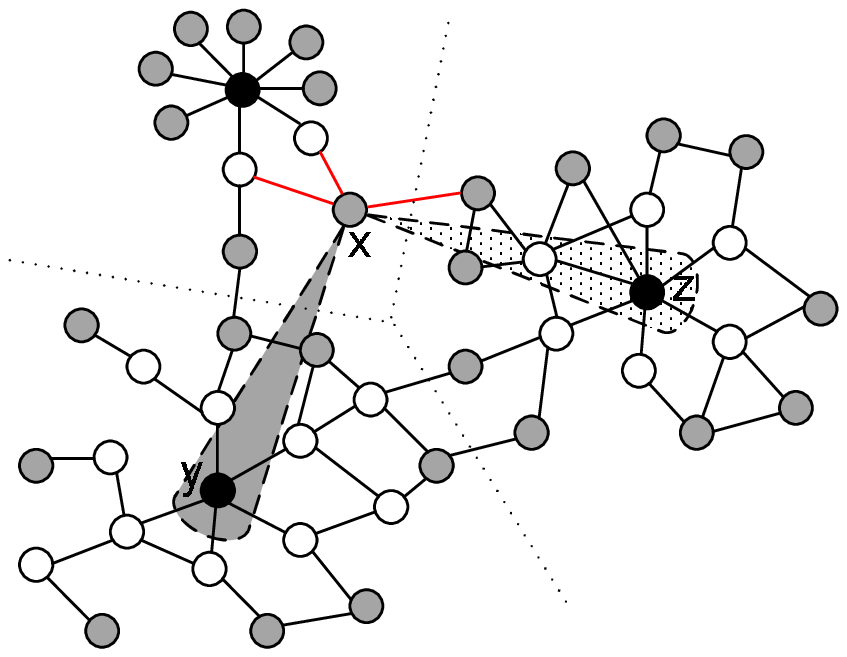}
            \label{subfig:fig2}
        }\quad
        \subfigure[Three unconnected components with three regions when $g=3$. Here, the node $x$ can create the beam either towards $y$ or $z$, but because the $hopcount$ to $y$ and $z$ is same as $\infty$, $x$ randomly decides between $y$ and $z$ to connect to.]
        {
            \includegraphics[width = 0.31\textwidth]{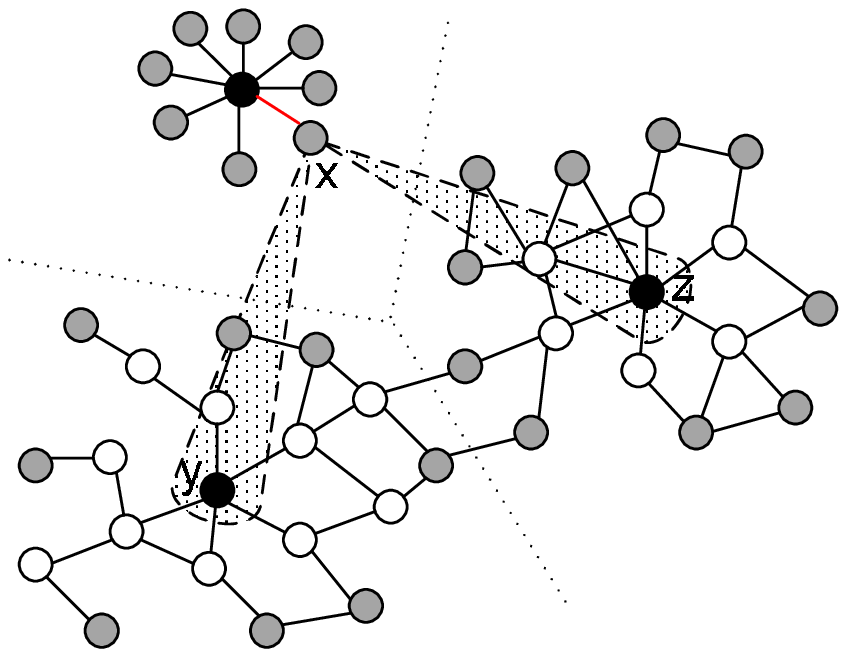}
            \label{subfig:fig3}
        }
        \subfigure
        {
            \raisebox{2.1cm}
            {
                \includegraphics[width = 0.27\textwidth]{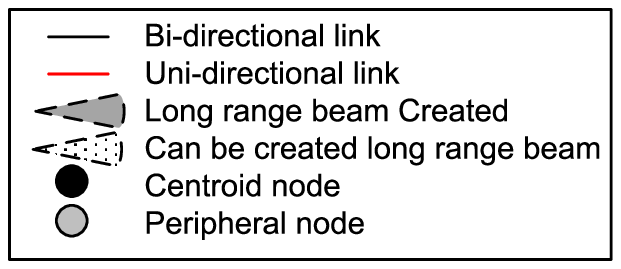}
            }
            \label{subfig:fig4}
        }
    }
    \caption{Beamforming priority.}
    \label{fig:f3}
\end{figure*}

To account for choosing the correct centroid to connect, the peripheral node, $\wp_{i}$, builds $RC_{\wp_{i}}^{*}$, a set of all centroid nodes reachable when it is beamforming. To determine $RC_{\wp_{i}}^{*}$, the peripheral nodes sweep through all the sectors ($m^2$) created with the chosen number of antenna elements except the sectors in which $\varrho_{\wp_{i}}$'s have created the beam. If $RC_{\wp_{i}}^{*}-RC_{\wp_{i}}\neq\emptyset$ and $|RC_{\wp_{i}}^{*}-RC_{\wp_{i}}|>1$, i.e., $\wp_{i}$ identified two or more potential centroid nodes, assuming the $hopcount$ to these centroid nodes as $\infty$ the decision to connect to one of them is randomly made. However, if $RC_{\wp_{i}}^{*}-RC_{\wp_{i}}=\emptyset$, i.e., no new centroid is found, the $\wp_{i}$ decides to connect to farthest centroid node in $RC_{\wp_{i}}$. As we know that $APL$ is dependent on $\sum_{v,w\in V,v\neq w}^{} hops(v,w)$ any reduction in this summation will lead to a reduced network path length. In order to have maximum reduction in the path length, the node should connect to the farthest centroid. If the farthest centroid node was the $c_{i}$, then $\wp_{i}$ beamforms towards it. However, this decision also depends on the $hopcount$ between $c_{i}$ and $\wp_{i}$. Creating the beam toward the centroid that is less than two hops away will only reduce the initial neighborhood but not the $APL$. In this case $\wp_{i}$ drops the decision of being the peripheral node and remains omnidirectional. The Fig. \ref{fig:f3}(a) and the Fig. \ref{fig:f3}(b) depicts the same. In the Fig. \ref{fig:f3}(a), node $x$ is 5 hops away from $y$ while it is 4 hops away from $z$ and 2 hops away from the centroid of the region in which $x$ lies. Thus, in order to have a reduced path length, node $x$ decides to create beam towards $y$. On the contrary, in the case when the node $x$ does not have the previously stored information about the centroid nodes $y$ and $z$, the node considers $hopcount$ to these centroid nodes as $\infty$ and randomly chooses one of them to connect to, (Cf. Fig. \ref{fig:f3}(b)).

\newsavebox{\tempbox}

\begin{figure}[ht]
    \hspace{-0.5cm}
    \includegraphics[width = \columnwidth]{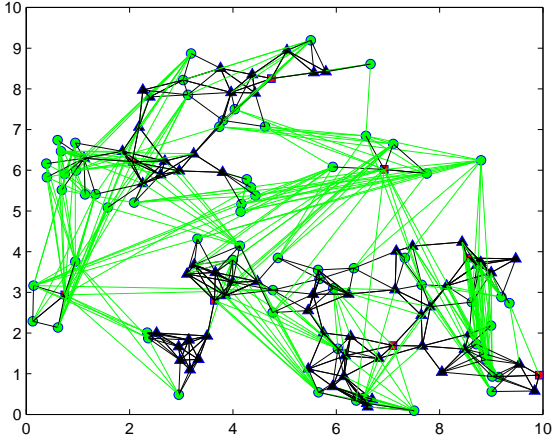}
    \caption{Nodes beamforming towards other region's centroid created for the Fig. \ref{subfig:fignodedistbnonunif} using $g<5$. The nodes marked in green beamform. The directional beams are also shown in green. The nodes marked with black triangle do not beamform. The nodes marked with red square are the centroid nodes.}
    \label{fig:beamform}
\end{figure}

Whenever a peripheral node creates a beam towards a centroid node that is more than 1 hop away, asymmetric link may arise. This is due to the fact that the $B_{l}$ of peripheral node is $m*r$ while $B_{l}$ of a centroid node is $r$, in other words, $\frac{B_l \; of \; Centoid}{B_l \; of \; peripheral}=\frac{1}{m}$. Due to this difference, peripheral nodes will not know if they got connected to the centroid of other region or not. We propose to solve this issue as, when a centroid node receives information about the node trying to connect to it, it just for one time instant, to acknowledge the reception, creates the beam back to the node. We do this after determining angle of incidence of the beam. This works well for both connected and unconnected components. Algorithm \ref{Algo2} represents a brief algorithmic description of beamforming using Flocking rule analogy. The Fig. \ref{fig:beamform} shows the new network created after running our algorithm on the network shown in the Fig. \ref{subfig:fignodedistbnonunif}.

\begin{algorithm}[!ht]
    \caption{Beamforming using Flocking Analogy}\label{Algo2}
    \begin{algorithmic}[1]
        \STATE $\backslash\backslash$ Alignment;
        \FORALL {\ensuremath{v_{i} \in V_{i} \in V}}
            \IF {\ensuremath{hops(v_{i},c_{i})>hops(L_{v,i},c_{i})}}
                \STATE \ensuremath{P_{i}=P_{i}+\{v_{i}\}}
                \STATE \ensuremath{P=P+\{v_{i}\}}
            \ENDIF
        \ENDFOR
        \STATE $\backslash\backslash$ Separation;
        \FORALL {\ensuremath{\wp_{i} \in P_{i} \in P}}
            \STATE set \ensuremath{m}
            \FORALL {$m^2$ Sectors $|\varrho_{\wp_{i}}(B_{b})$ \ensuremath{\notin} Sectors}
                \STATE \ensuremath{RC_{\wp_{i}}^{*} = RC_{\wp_{i}}^{*}+\{reachable\; centroid\; nodes\}}
            \ENDFOR
        \ENDFOR
        \STATE $\backslash\backslash$ Cohesion;
            \FORALL {\ensuremath{\wp_{i} \in P_{i} \in P}}
            \IF {\ensuremath{RC_{\wp_{i}}^{*}-RC_{\wp_{i}}\neq\emptyset}}
                \FORALL {\ensuremath{c \in RC_{\wp_{i}}^{*}-RC_{\wp_{i}}}}
                    \STATE \ensuremath{h = h+{hops(\wp_{i},c)}}
            \ENDFOR
        \ELSE
            \IF {\ensuremath{RC_{\wp_{i}}\neq\emptyset}}
                \FORALL {\ensuremath{c \in RC_{\wp_{i}}}}
                    \STATE \ensuremath{h = h+{hops(\wp_{i},c)}}
                \ENDFOR
            \ELSE
                \STATE \ensuremath{P_{i}=P_{i}-\{\wp_{i}\}}
            \ENDIF
        \ENDIF
        \STATE \ensuremath{beamtonode=\max\{h\}}
        \STATE \ensuremath{\theta} = Sector containing \ensuremath{beamtonode}
        \ENDFOR
    \end{algorithmic}
\end{algorithm}

\section {Formal Definitions}\label{sec:sec5}

\newtheorem{mydef}{Definition}
\newtheorem{Lemma1}{Lemma}

\begin{mydef}\label{defcentroid}
Assume a centroid $c_{i}$ of the region $G_{i}$, and a node $v_{c}$ in $V_{i}$ which has the highest Closeness Centrality, then
    \begin{eqnarray}\label{eq:remnodes}
        Closeness(v_{c}) &=& \operatorname*{arg\,\bf{max}}_{\forall v_{i} \in V_{i}} \left[Closeness(v_{i})\right] \nonumber\\
        Closeness(c_{i}) &\simeq& Closeness(v_{c})
    \end{eqnarray}
\end{mydef}

\begin{mydef}\label{defperiph}
    The node $v_{i}$ with neighborhood $L_{v,i}$ of the region $G_{i}$ with centroid $c_{i}$ is a peripheral node $\iff hops(v_{i},c_{i}) \geq hops(L_{v,i},c_{i})$.
\end{mydef}

\begin{Lemma1}\label{defremainnodes}
    The expected number of nodes remaining after applying the thinning processes, \cite{BettstetterGyarmati}, on a uniformly distributed network is

    \begin{equation}\label{eq:remnodes}
        E(n) = \rho A \left( 1-\frac{\Gamma (r_b,\rho r_b^{2}\pi)}{(r_b-1)!}\right)
    \end{equation}
    where $E(n)$ is the expected number of nodes remaining after the thinning process is applied, $\rho$ is the initial node density in a given area $A$ and $\Gamma(r_b,\rho r_b^{2}\pi)$ is the incomplete gamma function.
\end{Lemma1}

\begin{Lemma1}\label{defdistcnode}
    The separation between any two head nodes is between $(g, 2g+1]$ where $g$ is the $hopcount$ used to create the region, \cite{NagpalMamei}.
\end{Lemma1}
\begin{proof}\label{proofdefdistcnode}
    Consider a head node with a gradient $g$ around itself. All the nodes within $g$ hops from the head node will be in its region. A node which is more than $g$ hops away will lie in another region. If in the neighboring region, a head node does not have any gradient around it, then the distance between the two head nodes in hops will be $g+1$. On the other hand, if the neighboring region also has a gradient $g$ around it, then the distance between two head nodes in hops will be $2g+1$.
\end{proof}

\begin{Lemma1}\label{lemone}
    The number of regions is equal to number of centroid nodes and each region has exactly one centroid node.
\end{Lemma1}
\begin{proof}\label{proofone}
    Our algorithm computes the centroid of the region based on average of coordinates, Degree and Egocentric Betweenness of the node for each region. According to our algorithm, the nodes are termed as centroid if the node falls within $\varepsilon$ range of the centroid coordinate estimation algorithm and have maximum sum of Degree and Egocentric Betweenness. If still there are multiple nodes that are termed as centroid nodes, the nodes randomly decide for being the centroid and thus only one node is chosen as centroid. The value of $\varepsilon$ is thus an important factor in the estimation of the centroid node. Also, smaller $\varepsilon$ will tend to provide better estimation of the centroid nodes. As there is only one centroid node per region, the number of centroid nodes is equal to the number of regions.
\end{proof}

\begin{Lemma1}\label{lemnotcent}
    If a node is not a centroid node, it is connected to a centroid node.
\end{Lemma1}
\begin{proof}\label{prooflemnotcent}
     Our algorithm identifies regions and their centroid nodes. An identified region is always connected, i.e., all the nodes in the identified region are connected to each other. Further, there is one and only one centroid node in a region, ref. lemma \ref{lemone}. Thus for a given region, all nodes that are not centroid are connected to the centroid node.
\end{proof}

\begin{Lemma1}\label{lemuncon}
    An unconnected node is both the centroid node as well as the peripheral node.
\end{Lemma1}
\begin{proof}\label{prooflemuncon}
    A single unconnected node does not have any neighborhood. It thus remains uninhibited at the end of the region formation phase and becomes the head. As it is lacking any neighborhood, the node does not have any gradient around itself and is the only node in the region. In this region, the average coordinates perfectly match the virtual coordinates of the node. Thus requiring no further computation to correctly identify the centroid node.

    This node is also the peripheral node as the condition of Definition \ref{defperiph} holds true because of the unavailability of the neighborhood.
\end{proof}

\begin{Lemma1}\label{numregions}
    For a node distribution and fully connected network with average node density $\rho$ and total number of nodes $|V|$, then $|C|$ is bounded by $\frac{|V|}{\rho g^2r^2\pi}$ and $\frac{|V|}{\rho g^2r^2\sqrt{3}}$.
\end{Lemma1}
\begin{proof}\label{proofnumregions}
    From lemma \ref{defdistcnode}, the hop distance between two heads is bounded by $(g,2g+1]$.

    \textbf{\emph{Case 1 (Lower Bound)}}:
        When the heads are separated by $2g+1$ hops, the number of regions formed are less. The number of nodes in one region is $\rho g^2r^2\pi$. Thus, the total number of nodes in all the $N$ regions is $|N|\rho g^2r^2\pi$. As the total number of nodes are $|V|$, $\therefore |N|=\frac{|V|}{\rho g^2r^2\pi}$. From lemma \ref{lemone} $|C|=|N|$, $\therefore |C|=\frac{|V|}{\rho g^2r^2\pi}$

    \textbf{\emph{Case 2 (Upper Bound)}}:
        When all the heads are separated by $g+1$ hops, the number of regions formed are more. A head in such a case is connected to only 6 other heads. This can be visualized as a hexagon with vertex-vertex distance equal to $g+1$ and a node at the center of hexagon. Each of the vertex nodes are shared between 3 other hexagons. Thus, the total number of heads that are exclusive for the hexagon are $\frac{6}{3}$+1 = 3. In other words, there are 3 heads in an area of $\frac{6g^2r^2\sqrt{3}}{2}$. Thus, for the area=$\frac{|V|}{\rho}$, $|C|$=$\frac{|V|}{\rho g^2r^2\sqrt{3}}$.
\end{proof}

\begin{figure}[ht]
    \centering
    \includegraphics[width=\columnwidth]{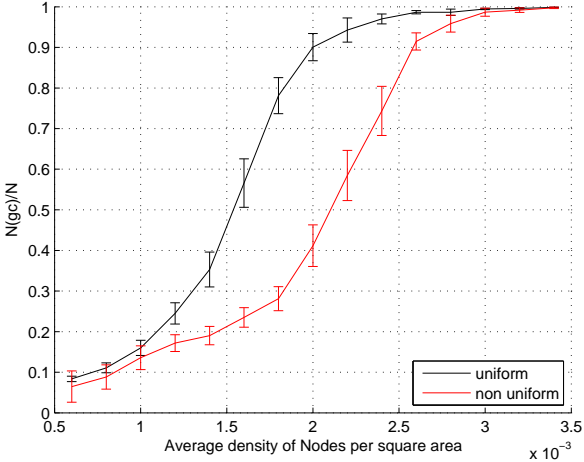}
    \caption{Percolation of the giant component for nodes distributed uniformly and non-uniformly. We use $r_b=30m$ and $\ell_{min}=5$ to achieve non-uniform distribution of nodes. There is a difference in the values of the size of the giant component at the same average density because the algorithm used to generate non-uniformity \cite{BettstetterGyarmati} tends to create clusters of nodes that might be unconnected. This leads to a network that is less connected than the uniformly deployed network. However, when the density increases the size of the clusters also increases.}
    \label{fig:percolation}
\end{figure}

\begin{Lemma1}\label{numregionsunconnected}
    Consider a network with $j$ components ($j>1$), average density of the nodes as $\rho_{k}$ and number of nodes as $|V_{k}^{j}|$ for $k\in j$, $|C|$ is bounded by $\sum_{k=1}^{j} \frac{|V_{k}^{j}|}{\rho_{k} g^2r^2\pi}$ and $|V|$, where $|V|=\sum_{k=1}^{j} |V_{k}^{j}|$.
\end{Lemma1}
\begin{proof}\label{proofnumregionsunconnected}
    From lemma \ref{defdistcnode}, the hop distance between two heads is bounded by $(g,2g+1]$.

    \textbf{\emph{Case 1 (Lower Bound)}}:
        Consider $k^{th}$ component of the network. When the heads are separated by $2g+1$ hops, the number of regions formed is less. The number of nodes in one region is $\rho_{k} g^2r^2\pi$. Thus, the total number of nodes in all the regions in the component is $N_{k}\rho_{k} g^2r^2\pi$, where $N_{k}$ are the number of region in $k^{th}$ component. But as the total number of nodes were assumed to be $|V_{k}^{j}|$, $\therefore$ $N_{k}$=$\frac{|V_{k}^{j}|}{\rho_{k} g^2r^2\pi}$. Thus for all the components, the number of regions formed is $|N|=\sum_{k=1}^{j}N_{i} = \sum_{k=1}^{j} \frac{|V_{k}^{j}|}{\rho_{k} g^2r^2\pi}$. From lemma \ref{lemone}, $|C|=|N|$, $\therefore |C|=\sum_{k=1}^{j} \frac{|V_{k}^{j}|}{\rho_{k} g^2r^2\pi}$

    \textbf{\emph{Case 2 (Upper Bound)}}:
        Upper bound to the number of regions arises when all nodes in the network are disconnected. Thus, all nodes in such a case will be uninhibited thereby becoming region heads. Thus $|C|=|V|$.
\end{proof}

\begin{figure}[ht]
    \centering
    \includegraphics[width=\columnwidth]{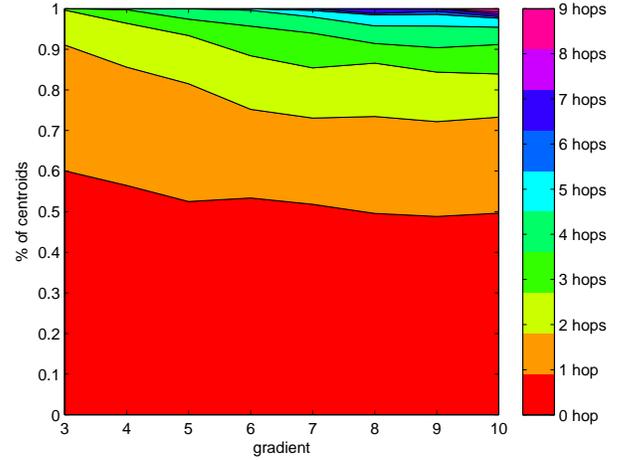}
    \caption{Relationship between centroid nodes and the nodes having maximum Socio-Centric Betweenness.}
    \label{fig:relation}
\end{figure}

\begin{Lemma1}\label{numpnodes}
    For a node distribution and fully connected network, and using lemma \ref{numregions}, the number of peripheral nodes in the network is bounded by $\frac{|V|(2g+1)}{g^2}$ and $\frac{|V|(2g+1)\pi}{g^2\sqrt{3}}$.
\end{Lemma1}
\begin{proof}\label{proofpnodes}
    Peripheral nodes are the nodes lying in the outer most gradient of the region. Thus, the number of nodes in the $g^{th}$ gradient of a region = $\rho g^2r^2\pi - \rho (g-1)^2r^2\pi$ = $\rho (2g+1)r^2\pi$

    Now using lemma \ref{numregions}, the number of peripheral nodes for all regions thus varies between $\frac{|V|(2g+1)}{g^2}$ and $\frac{|V|(2g+1)\pi}{g^2\sqrt{3}}$.
\end{proof}

\begin{Lemma1}\label{numpnodesdisconnected}
    For a node distribution and network with $j$ components ($j>1$), and using lemma \ref{numregionsunconnected} and lemma \ref{lemuncon}, the number of peripheral nodes in the network is bounded by $\sum_{k=1}^{j} \frac{|V_{k}^{j}|(2g+1)}{g^2}$ and $|V|$.
\end{Lemma1}
\begin{proof}\label{proofpnodesdisconnected}
    Peripheral nodes are the nodes lying in the outer most gradient of the region. Thus, the number of nodes in $g^{th}$ gradient of a region in $k^{th}$ component = $\rho_{k} g^2r^2\pi - \rho_{k} (g-1)^2r^2\pi$ = $\rho_{k} (2g+1)r^2\pi$.

    Now using lemma \ref{numregionsunconnected} and lemma \ref{lemuncon}, the number of peripheral nodes for all regions thus varies between $\sum_{k=1}^{j} \frac{|V_{k}^{j}|(2g+1)}{g^2}$ and $|V|$.
\end{proof}

\section{Simulation setup}\label{sec:sec6}

We use a simulation area of $A=500m$x$500m$ to simulate our algorithm. $r_b$ and $\ell_{min}$ are set to $30m$ and $5$ respectively to achieve the non-uniform distribution of node throughout the simulation area. The non-uniform node distribution enables us to visualize the real world scenarios. The range of average density, $\rho$, of nodes per unit area is set to [$1$x$10^{-3}$, $2.5$x$10^{-3}$]. We make the choice of this range for $\rho$ after considering the percolation of the giant component for the non-uniform node deployment, (Cf. Fig. \ref{fig:percolation}). Initially, each node operates in omnidirectional mode using $m=1$ antenna element with the omnidirectional radius as $r=30m$. We set the maximum number of antenna elements that the nodes are equipped with to $M=6$. The separation between two antenna elements computed using $WiFi$ frequency, $f=2.4Ghz$. Through our simulations, we explore the effect on connectivity, $APL$ and $CC$ by varying the node densities and the gradient.

We use MATLAB to simulate our algorithm with a confidence interval of $95\%$. We average All the results over $50$ topologies.

\begin{figure*}
    \centering
    \mbox
    {
        \subfigure[Average path length in hops.]
        {
            \includegraphics[width = \columnwidth]{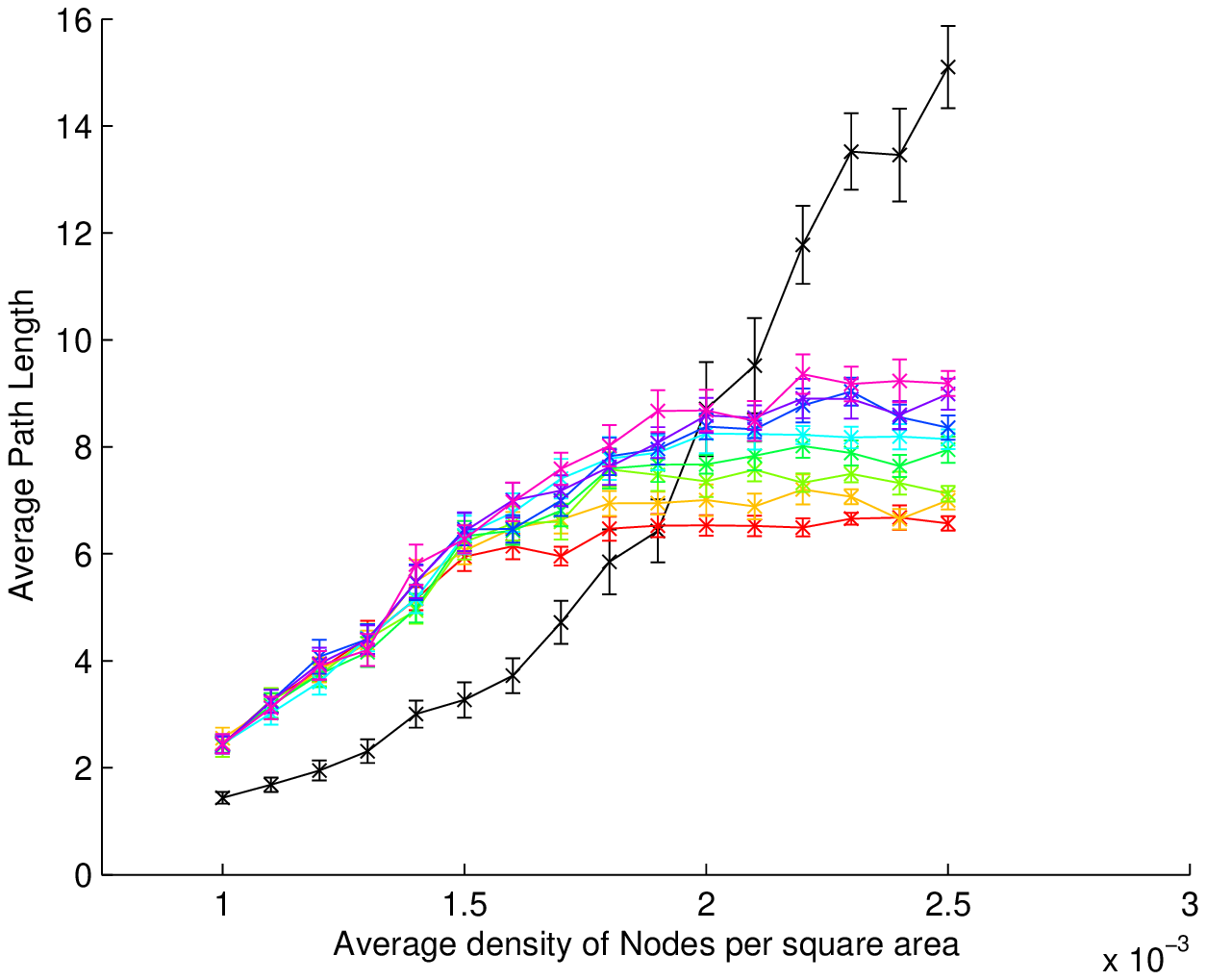}
            \label{subfig:avlsec}
        }\quad
        \subfigure[Clustering Coefficient.]
        {
            \includegraphics[width = \columnwidth]{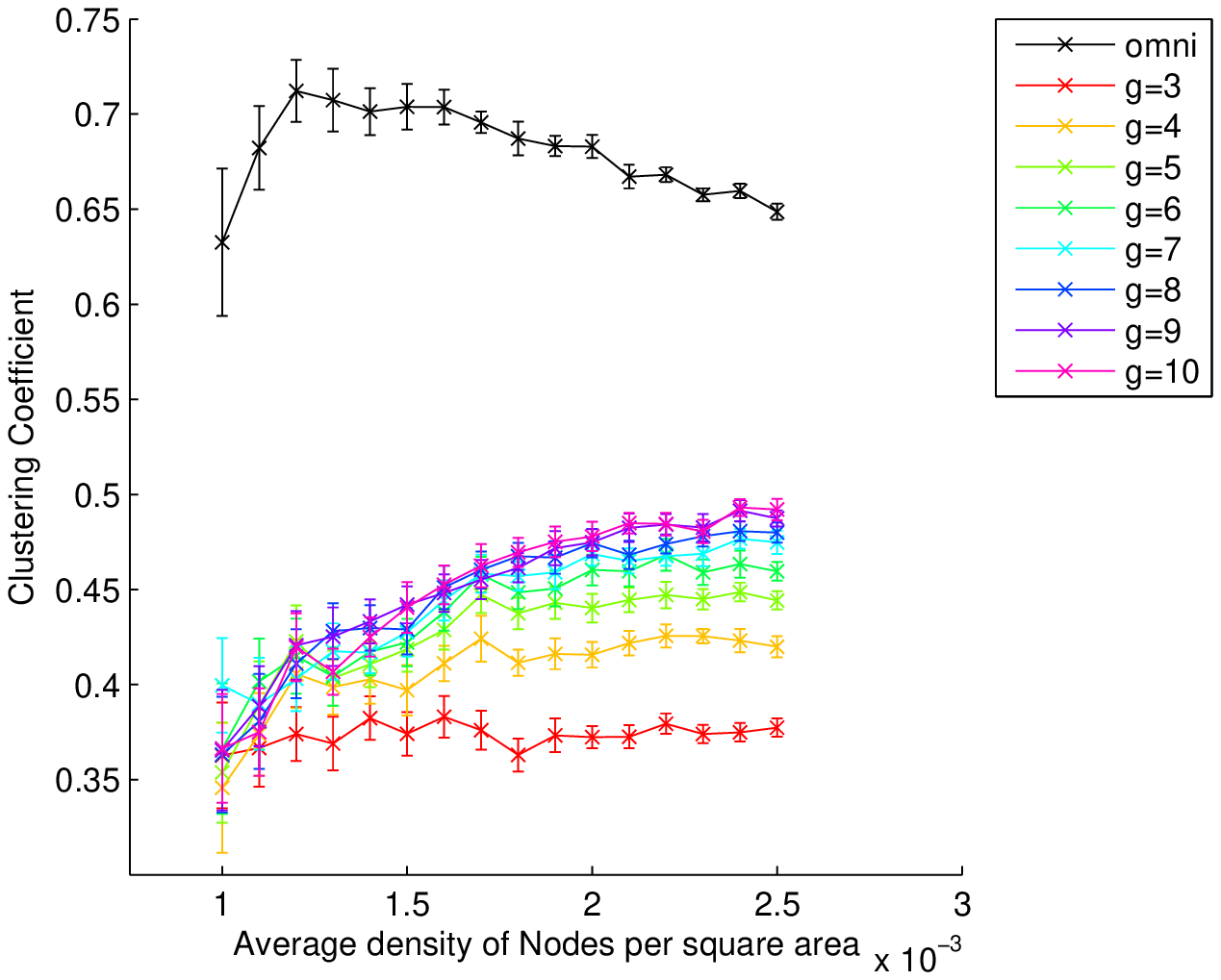}
            \label{subfig:ccsec}
        }
    }
    \mbox
    {
        \subfigure[Fraction of nodes designated as Peripheral nodes.]
        {
            \includegraphics[width = \columnwidth]{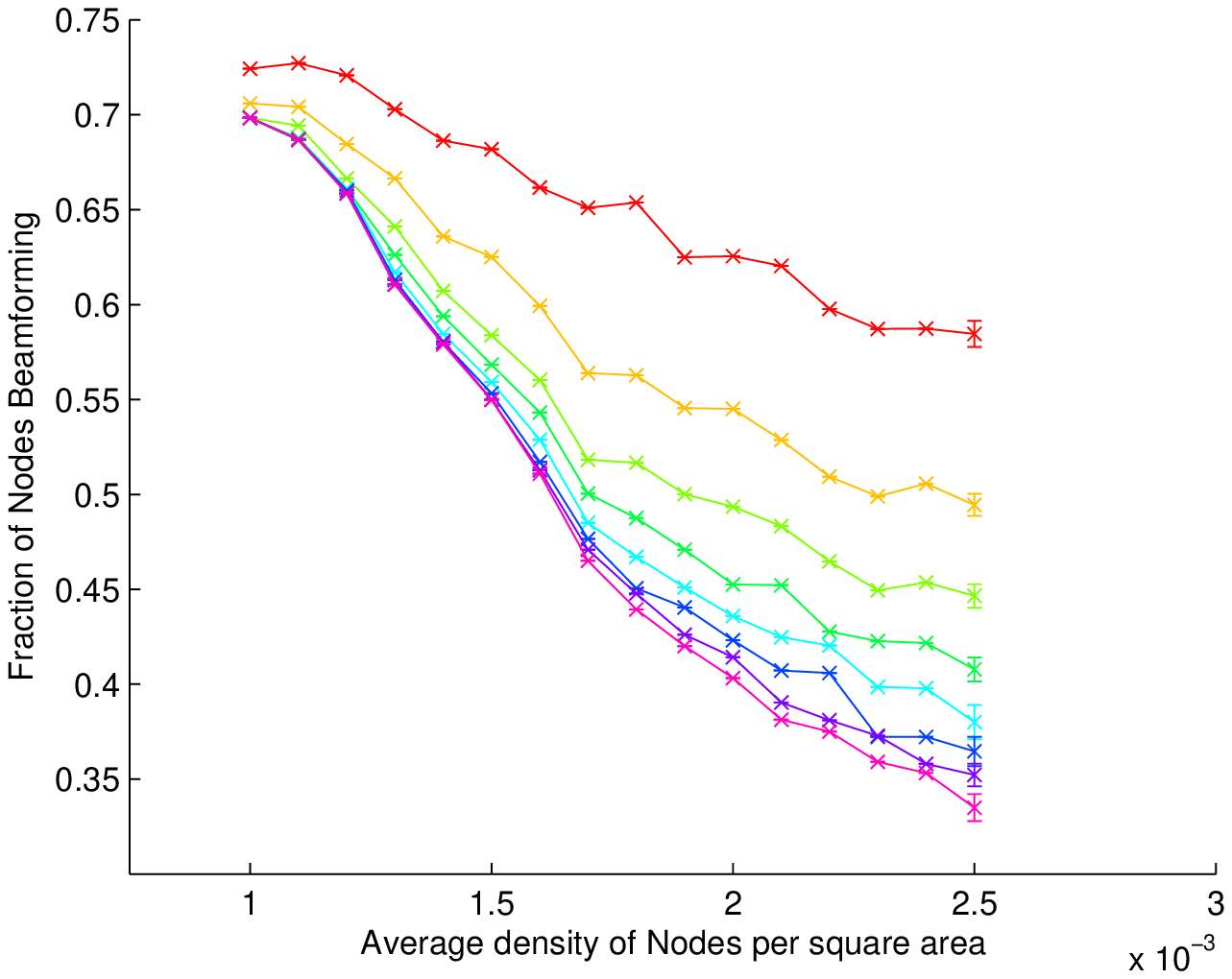}
            \label{subfig:pnodesec}
        }\quad
        \subfigure[Fraction of nodes designated as centroid nodes.]
        {
            \includegraphics[width = \columnwidth]{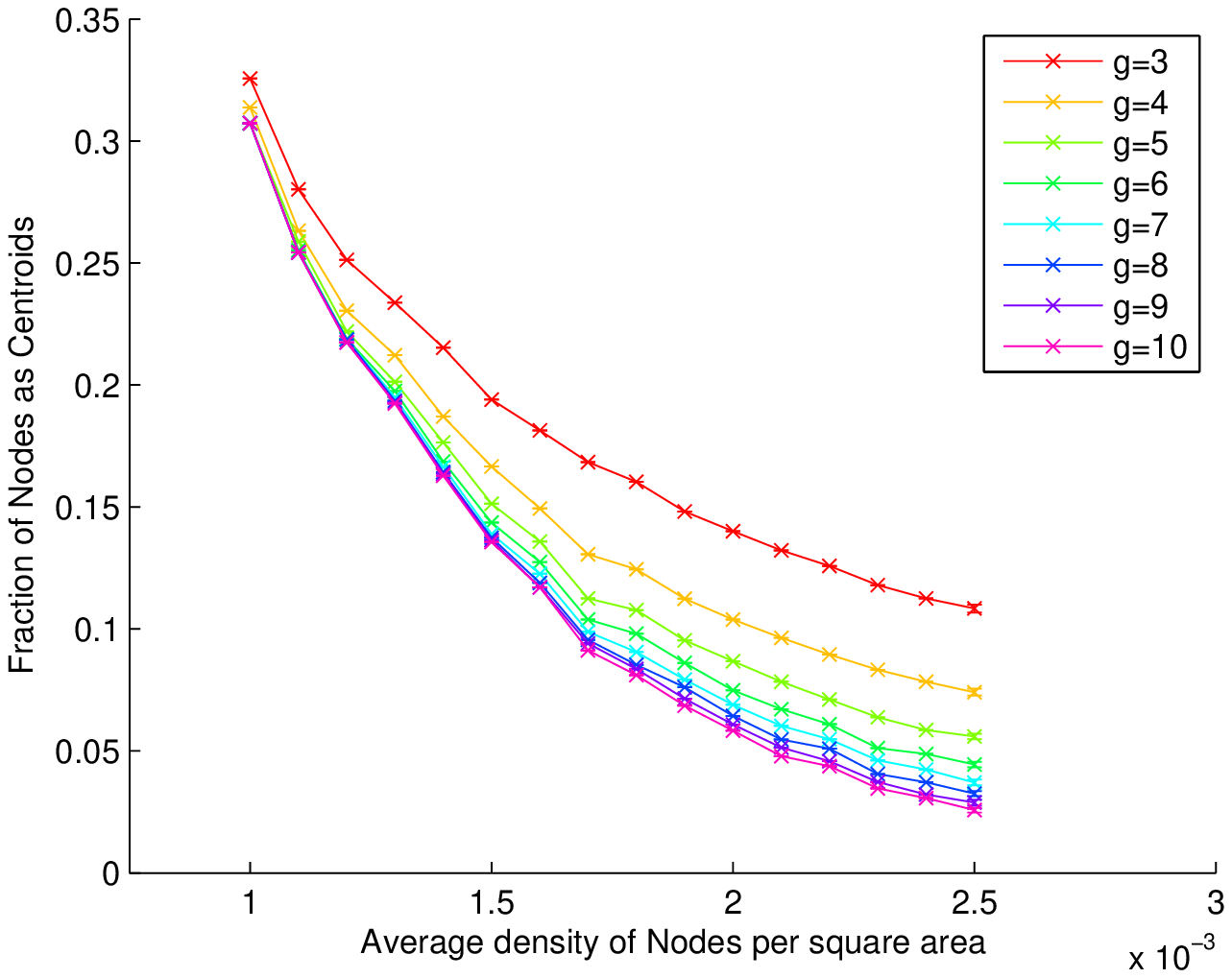}
            \label{subfig:cnodesec}
        }
    }
    \mbox
    {
        \subfigure[Number of components in the network.]
        {
            \includegraphics[width = \columnwidth]{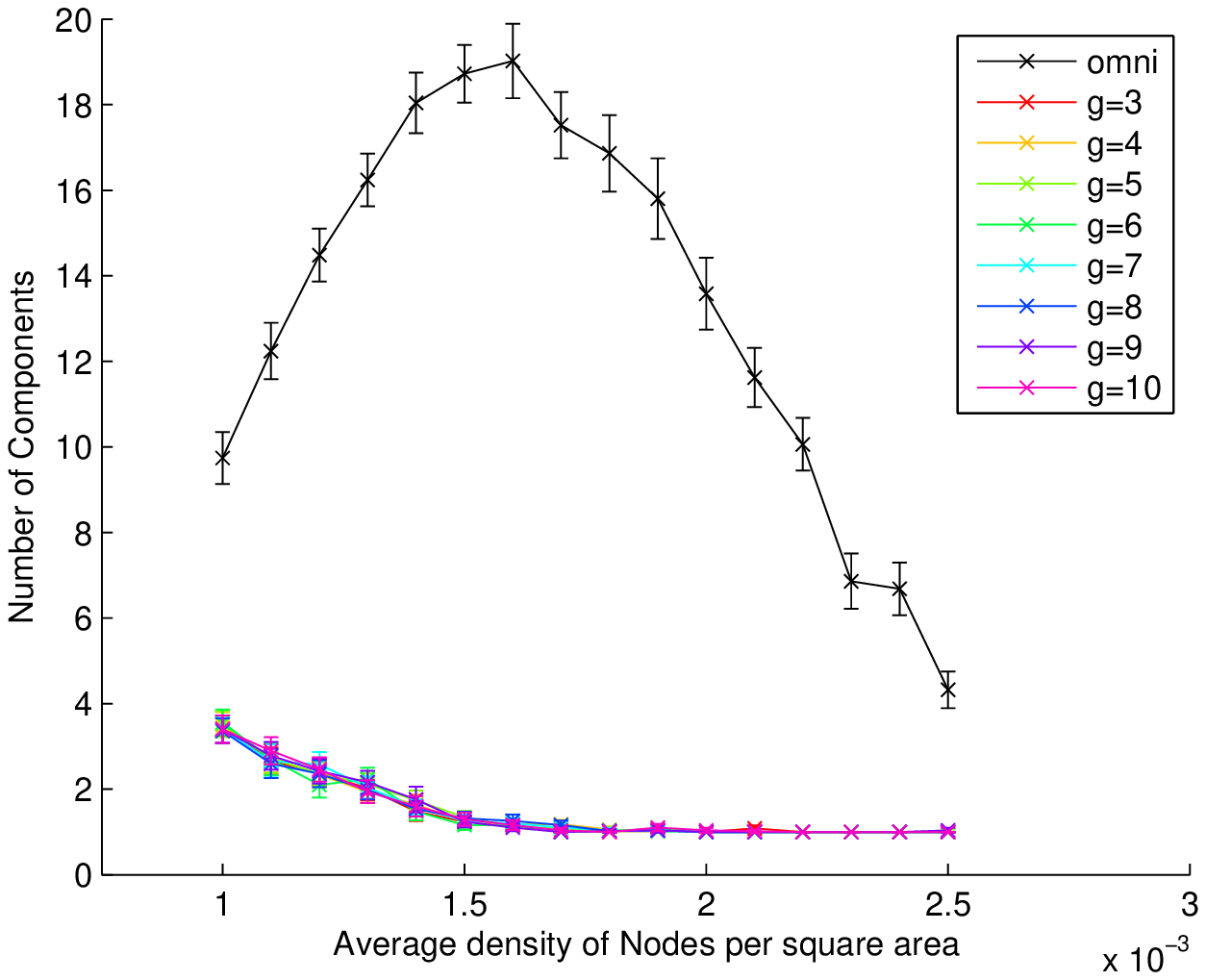}
            \label{subfig:connectivitysec}
        }\quad
        \subfigure[Normalized directional $APL$ and $CC$ for $N=625$ showing\newline the effects of the gradient.]
        {
            \includegraphics[width = \columnwidth]{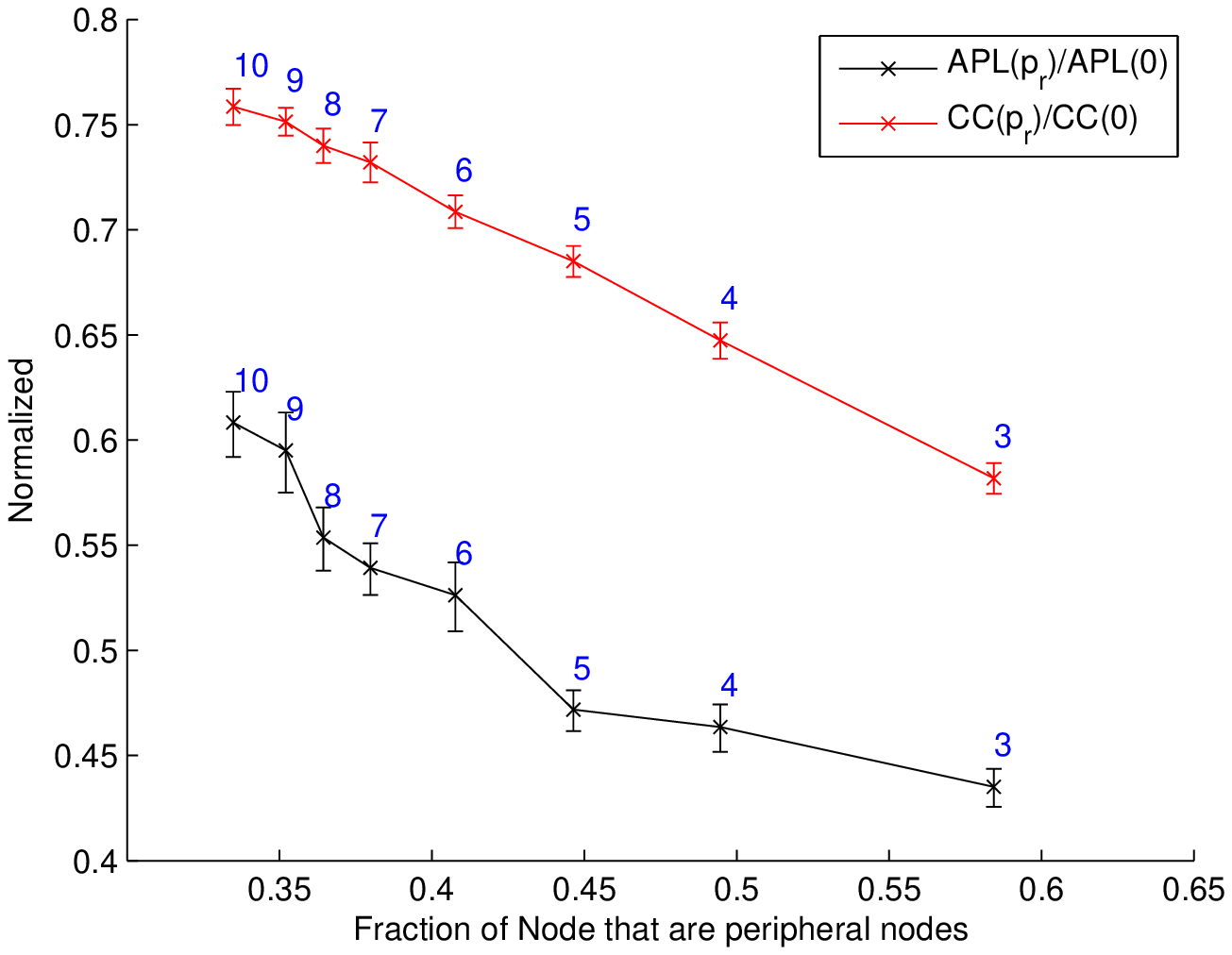}
            \label{subfig:smsec}
        }
    }
    \caption{Results obtained for $g\in[3,10]$, when we use Sector model and non-uniform node distribution.} \label{fig:unifbiobeamsec}
\end{figure*}

\begin{figure*}
    \centering
    \mbox
    {
        \subfigure[Average path length in hops.]
        {
            \includegraphics[width = \columnwidth]{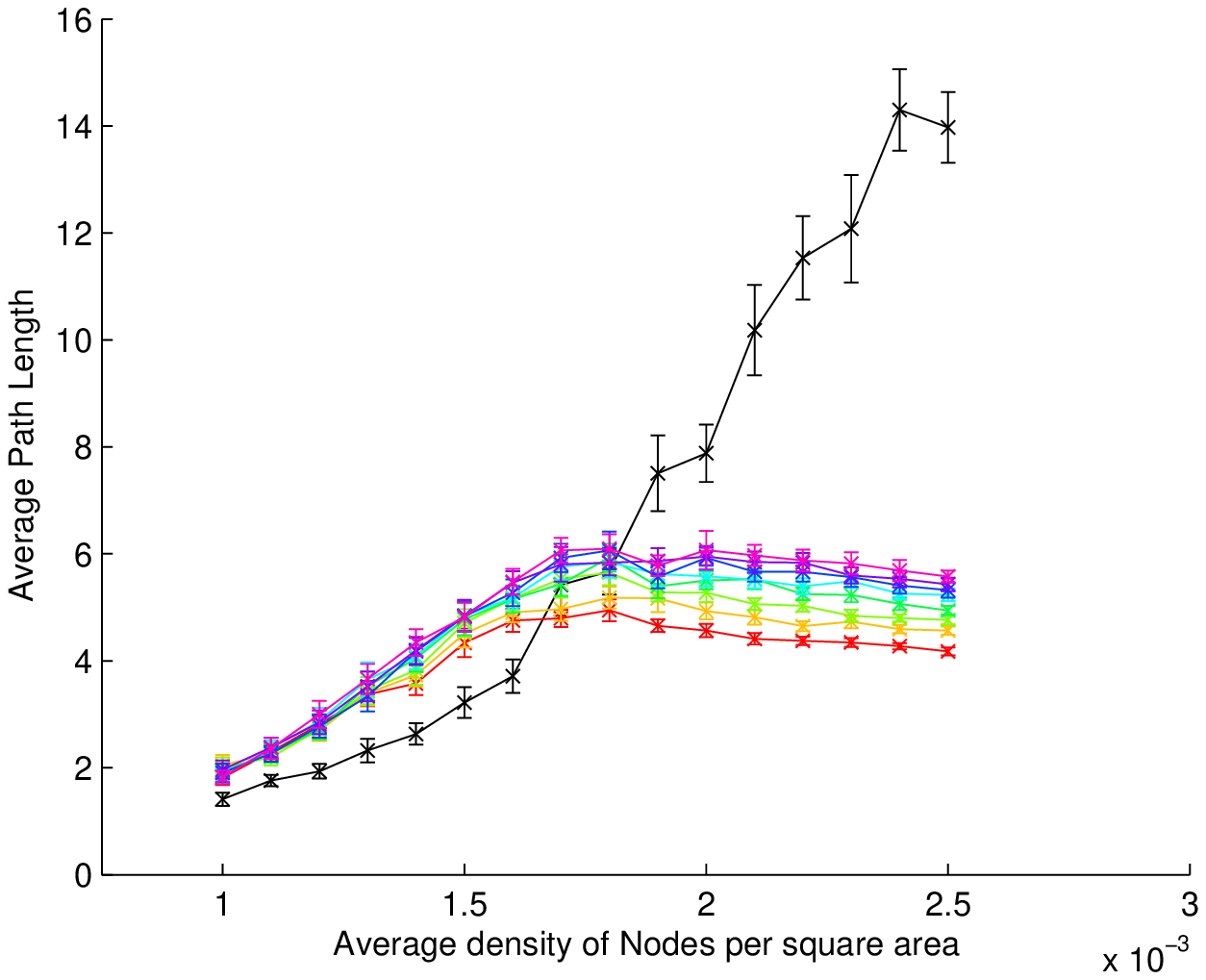}
            \label{subfig:avlULA}
        }\quad
        \subfigure[Clustering Coefficient.]
        {
            \includegraphics[width = \columnwidth]{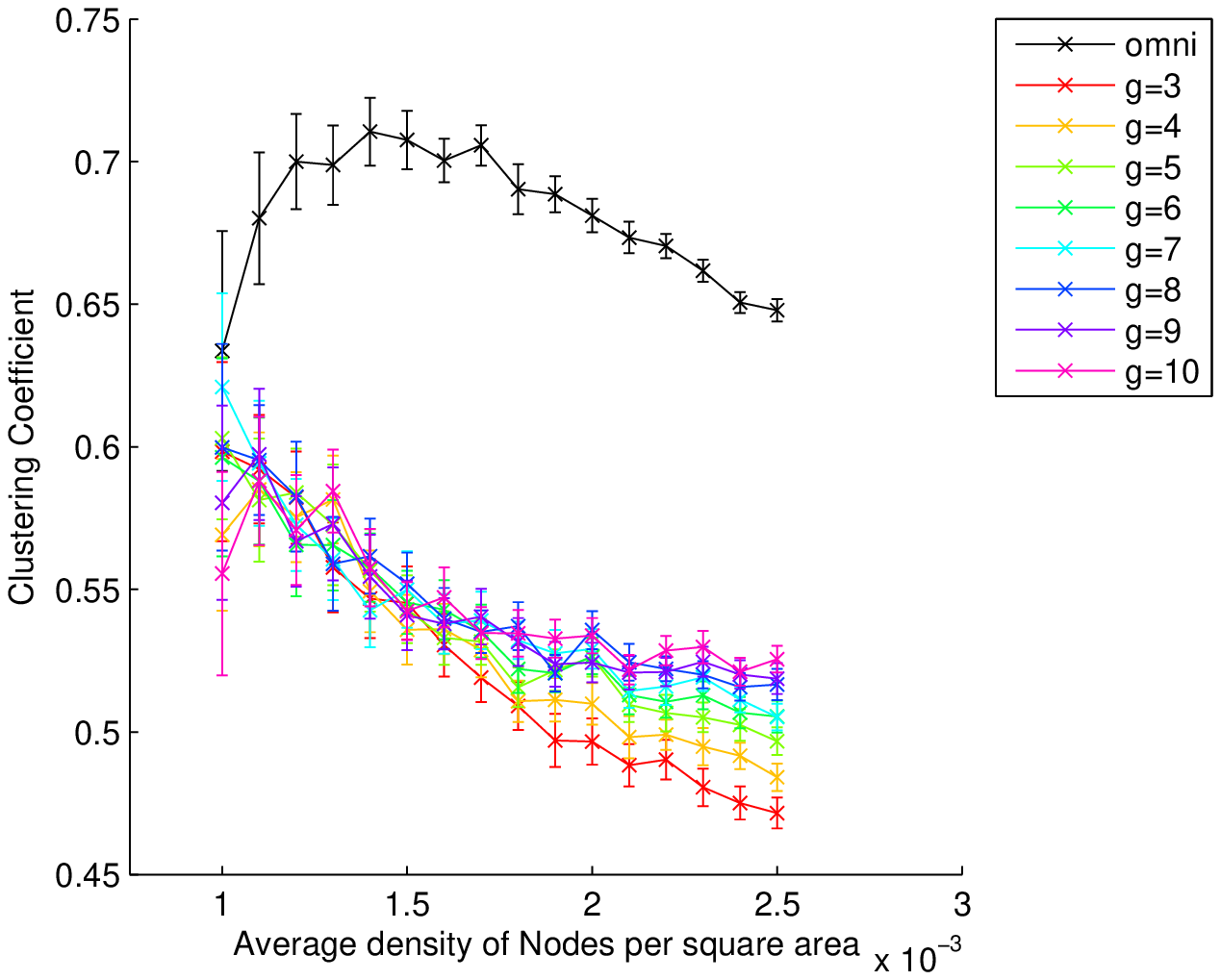}
            \label{subfig:ccULA}
        }
    }
    \mbox
    {
        \subfigure[Fraction of nodes designated as Peripheral nodes.]
        {
            \includegraphics[width = \columnwidth]{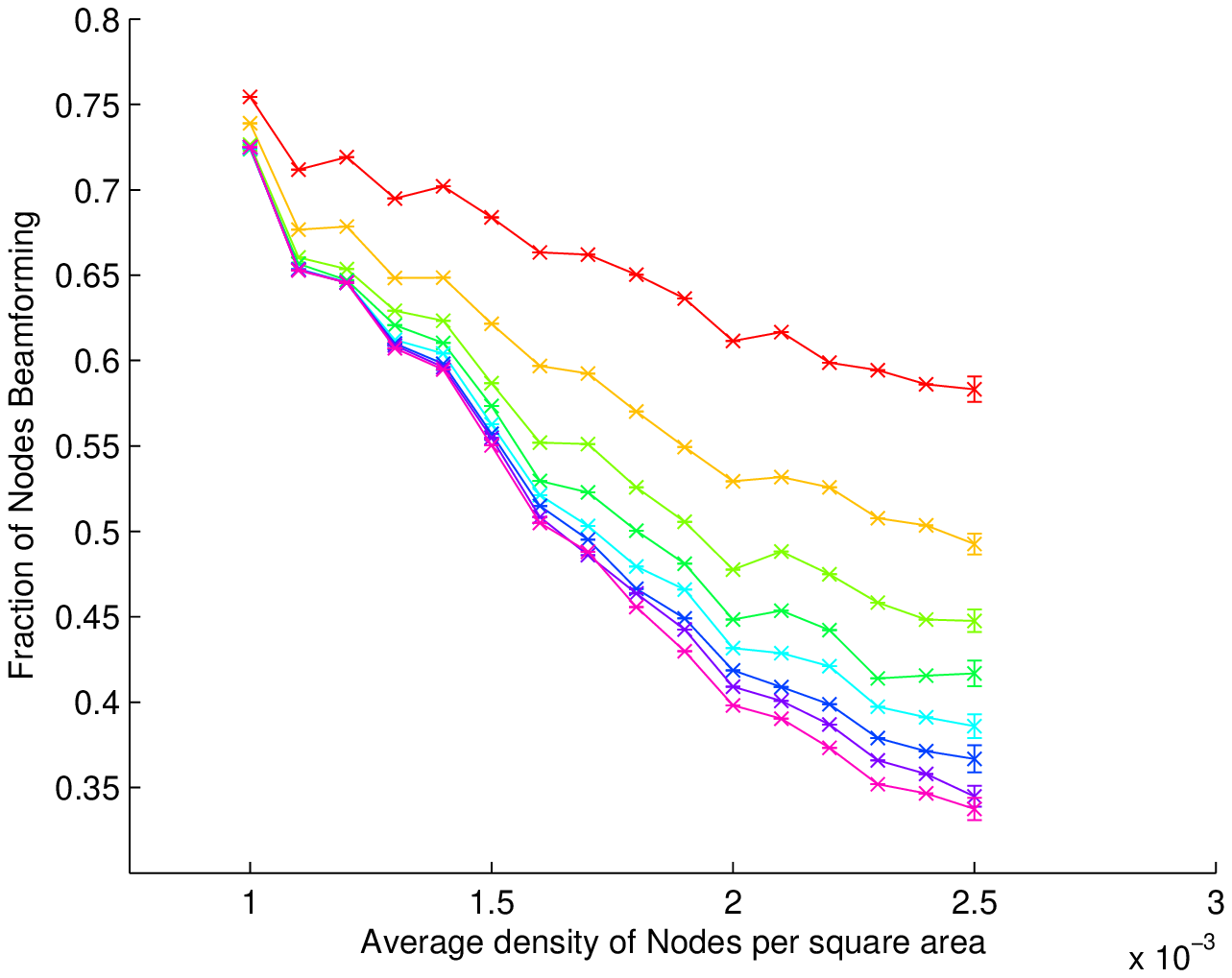}
            \label{subfig:pnodeULA}
        }\quad
        \subfigure[Fraction of nodes designated as centroid nodes.]
        {
            \includegraphics[width =\columnwidth]{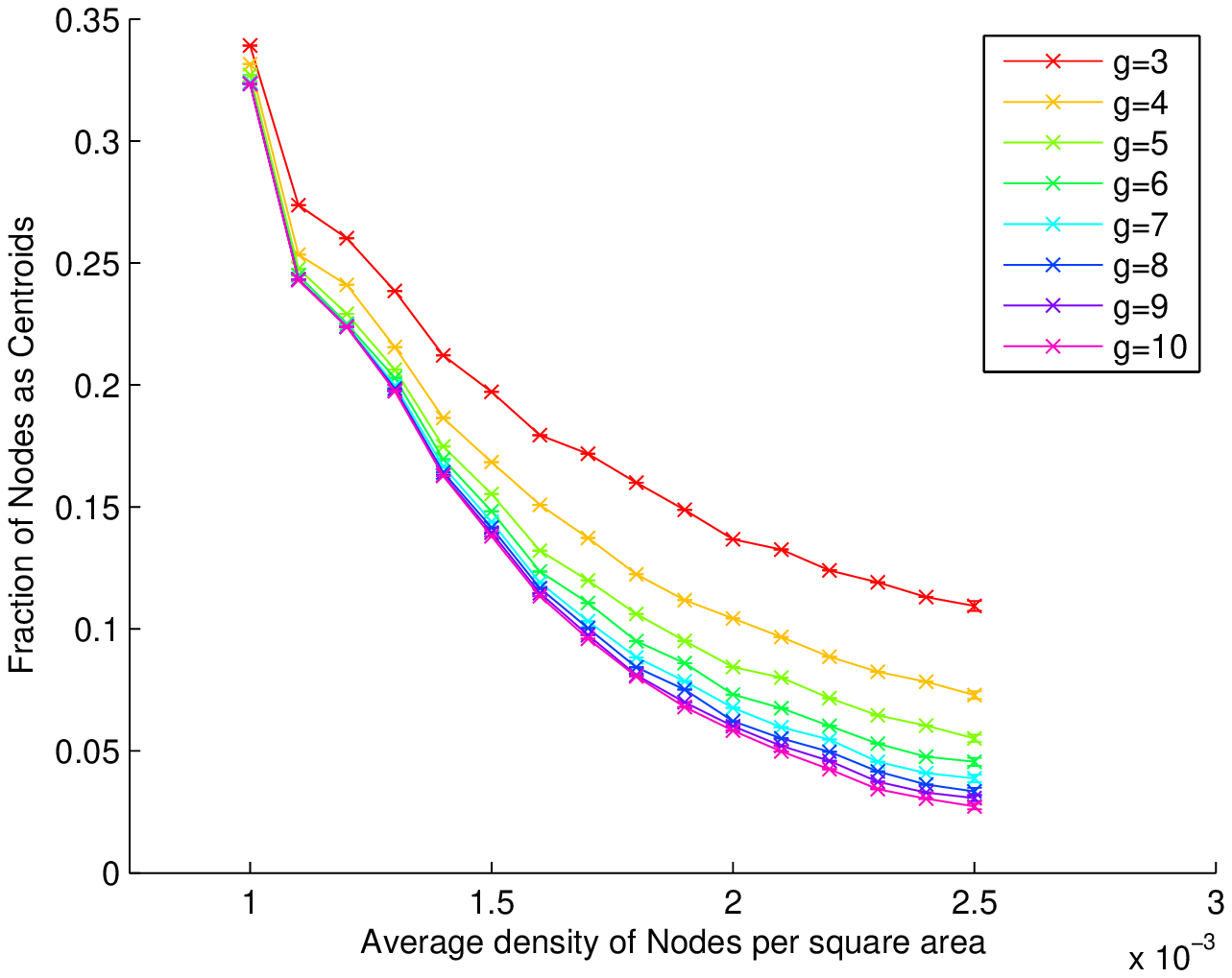}
            \label{subfig:cnodeULA}
        }
    }
    \mbox
    {
        \subfigure[Number of components in the network.]
        {
            \includegraphics[width = \columnwidth]{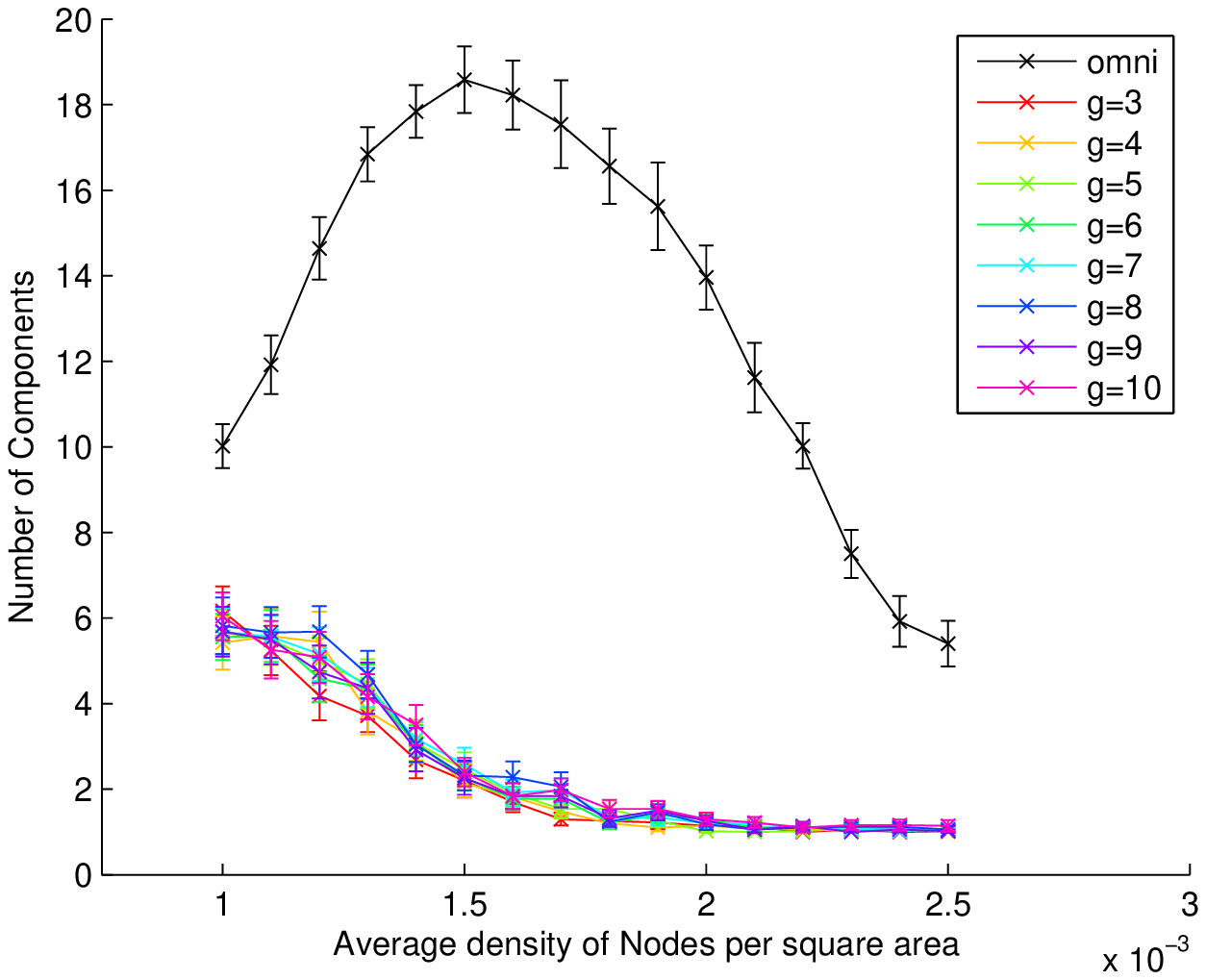}
            \label{subfig:connectivityULA}
        }\quad
        \subfigure[Normalized directional $APL$ and $CC$ for $N=625$ showing\newline the effects of the gradient.]
        {
           \includegraphics[width = \columnwidth]{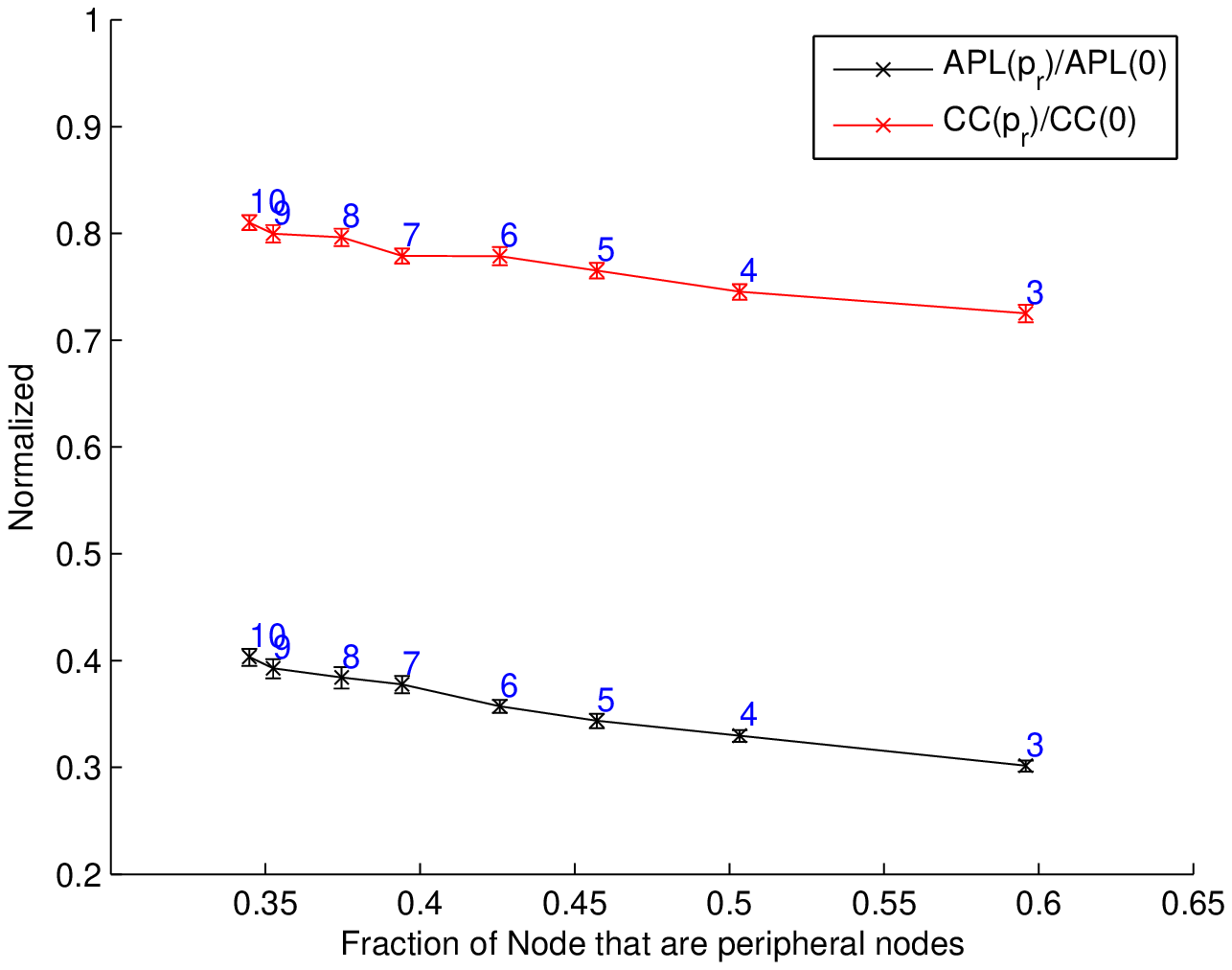}
            \label{subfig:smULA}
        }
    }
    \caption{Results obtained for different $g\in[3,10]$, using $ULA$ model and non-uniform node distribution.}
    \label{fig:unifbiobeamULA}
\end{figure*}

\section{Results and Analysis}\label{sec:sec7}

First, we prove the correctness of the centroid finding in the region. For this, we compute the relation between the nodes that have maximum Socio-Centric Betweenness and the centroid nodes in the region. If the centroid node has the highest Socio-Centric Betweenness in the region, then the algorithm found centroid node correctly, (Cf. Fig. \ref{fig:relation}). This depends on the value of the gradient. Larger gradients decrease the Socio-Centric Betweenness rank of the centroid node in the region. As the gradient increases, more nodes are now associated to a region thereby increasing the possibility of occurrences of the bridge nodes (bridge nodes have high Socio-Centric Betweenness value). Thus, we also calculate the distance in hops between the centroid node and the maximum Socio-Centric Betweenness node. According to the results, (Cf. Fig. \ref{fig:relation}), for a $g$, the percentage of centroid nodes that also have high Socio-Centric Betweenness is more and all the centroid nodes in the network are within $hopcount<g$. The Fig. \ref{fig:relation} however shows that for any $g\in[3,10]$ more than $95\%$ of the time the centroid node is within $4$ hop distance to the maximum Socio-Centric Betweenness value node and it is within $1$ hop $60\%$ of the time.

Further, we use $g\in[3,10]$ to obtain the results when the Sector model is used in a non-uniformly distributed network, (Cf. Fig. \ref{fig:unifbiobeamsec}). The Fig. \ref{subfig:avlsec} shows the effect of beamforming on the $APL$. The $APL$ obtained in omnidirectional case is initially less than that obtained for the directional cases because the density of the nodes in the component is low. When the algorithm induces directional beams, due to the inclusion of the nodes in other network components, there is an increase in the $APL$. The $APL$ for the directional case is less than that of the omnidirectional case when $\rho>2*10^{-3}$ due to the fact that the nodes connect to the centroid node of other regions in the different component as well as in the same component. The gradient affects the $APL$. The lower the value of the gradient is, higher is the number of nodes that beamform, (Cf. Fig. \ref{subfig:pnodesec}), leading to more shortcuts and in turn more reduction in the $APL$. For $\rho=2.5*10^{-3}$ and $g=10$, there is a reduction of almost 40\% in the $APL$ while there is a reduction of almost $55\%$ for $g=3$, (Cf. Fig. \ref{subfig:smsec}). However, for $\rho=1*10^{-3}$ and $g\in[3,10]$ when most nodes are unconnected, there is an increase of $70\%$ in $APL$ due to the above-mentioned facts.

The introduction of the long-range beams also causes the $CC$ to change, (Cf. Fig. \ref{subfig:ccsec}). For very low-density networks, the $CC$ for the directional case is less because beamforming leads to loss in the initial neighborhood. However, for higher density networks, the $CC$ does not vary as much as the $APL$ (Cf. Fig. \ref{subfig:smsec}). For $\rho=2.5*10^{-3}$, there is a reduction of $25\%$ and $38\%$ for $g=10$ and $g=3$ respectively. However, for $\rho=1*10^{-3}$ and any $g\in[3,10]$, the reduction in $CC$ is almost $40\%$. The $CC$ for directional case for $g<6$ and $\rho\in[1*10^{-3},2.5*10^{-3}]$ is almost constant. This implies that the directional network shows modularity where $CC$ is independent of $|V|$ and evolves towards hierarchical network \cite{Ravaszv}. However, when $g\in(6,10]$ the evolution towards hierarchical networks cannot be justified.

The number of components in the network can define connectivity. In a very low-density omnidirectional network, the number of disconnected components is higher, (Cf. Fig. \ref{subfig:connectivitysec}. The number of disconnected components increases to a certain maximum and then decreases as the density increases. This is because, for a high density, all nodes can find at least one neighborhood node within their reach. In addition, as the number of components decreases, the connectivity increases. For the directional case however, as nodes beamform towards different components with the objective of increasing connectivity, the number of disconnected components is less than that of the omnidirectional case.

\begin{figure}
    \centering
    \includegraphics[width=\columnwidth]{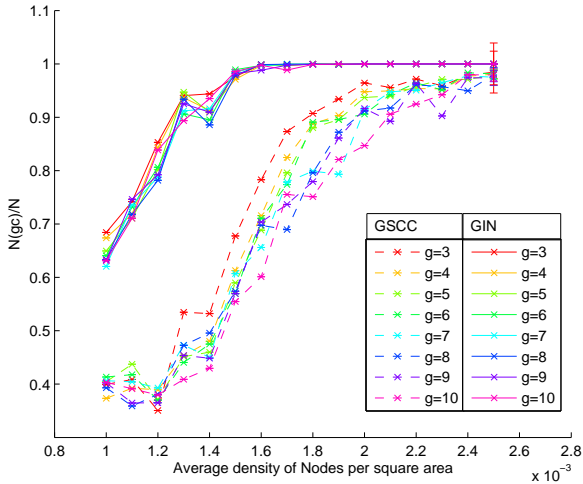}
    \caption{Variation in the size of the $GSCC$ and the $GIN$ for different density of nodes and $g \in [3,10]$.}
    \label{fig:percGSCCGIN}
\end{figure}

\begin{figure}
    \centering
    \includegraphics[width=\columnwidth]{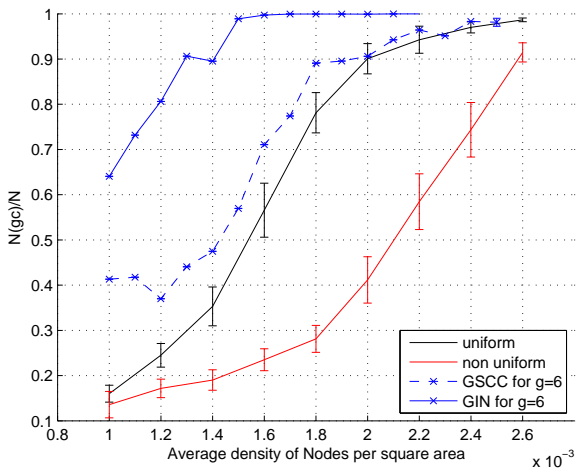}
    \caption{Comparison of size of the $GSCC$ and the $GIN$ for directed network with that of omnidirectional network for $g=6$.}
    \label{fig:perccomp}
\end{figure}

The size of the giant component can also explain the connectivity of the network. For the directed graphs however, \cite{Dorogovtsev} defined the giant component using the Giant Strongly Connected Component ($GSCC$)\footnote[3]{$GSCC$ in a directed graph is the length of the largest cycle in the graph component.} and the Giant In-Component ($GIN$)\footnote[4]{$GIN$ is the set of nodes in the component which can connect to $GSCC$.}. Thus, we calculate the size of $GSCC$ and $GIN$. We further show the difference between the size of the giant component for omnidirectional network, $GSCC$ and $GIN$. As stated in \cite{Dorogovtsev} that $GSCC \subset GIN$, we also observe that $GIN$ is a bigger set and contains more nodes than $GSCC$. $GIN$ reaches percolation very early, (Cf. Fig. \ref{fig:percGSCCGIN}). Comparing the size of the $GSCC$ of directional network with the giant component of the omnidirectional network, (Cf. Fig. \ref{fig:percolation}), we see that the size of $GSCC$ varies between [$0.84, 0.94$] for $\rho=2*10^{-3}$ for different values of the gradient while the size of giant component for the omnidirectional network is $0.41$. Thus, we observe an increase of almost $2.1$ times. The Fig. \ref{fig:perccomp} shows an increase of almost $2.2$ times when we compare of size of the $GSCC$ and the $GIN$ for $g=6$ with the giant component of the omnidirectional network.

The number of centroid nodes ($|C|$) depends on the value of the gradient, (Cf. Fig. \ref{subfig:cnodesec}). For a low-density network, the value of the gradient does not matter while as the density increases the value of the gradient affects the number of regions formed. As the gradient increases, more nodes inhibit leading to less number of regions. The difference between the number of regions formed for $g=3$ and $g=10$ is of $40$ for $\rho=2.5*10^{-3}$ while the difference for $\rho=1*10^{-3}$ is very less.

The value of the gradient used also affects the number of peripheral nodes ($|P|$) identified, (Cf. Fig. \ref{subfig:pnodesec}). For a low gradient value, as there are more regions, more nodes are included in $P$ because of the reduced neighborhood with respect to the region. However, when the value of the gradient is more, $|P|$ is less because there are more nodes in the region and the nodes have relatively more neighbors to check before making the decision of beamforming. $|P|$ greatly affects the number of unidirectional paths. However, it has an adverse effect on the $CC$. As the number of peripheral nodes increases, unidirectional paths between the nodes also increases leading to more loss in the $CC$. For $\rho=1*10^{-3}$ and $g\in[3,10]$, the difference between the number of peripheral nodes is almost negligible. For $\rho=2.5*10^{-3}$, however, the number of peripheral nodes varies by more than $120$ as the regions formed for lower gradient are more.

\begin{figure}
    \centering
    \includegraphics[width=\columnwidth]{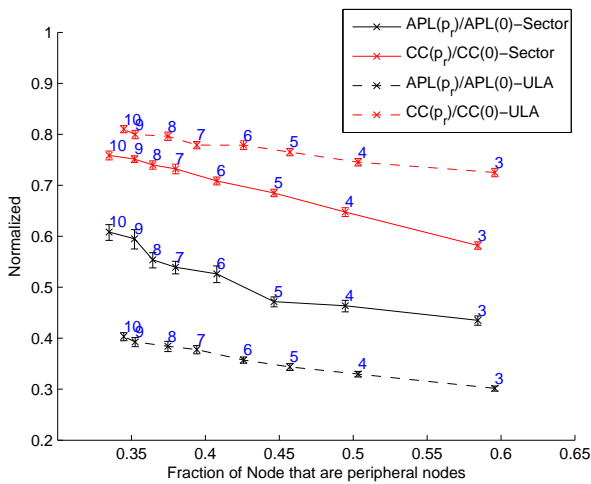}
    \caption{Normalized $APL$ and the $CC$ for $N=625$ showing the effects of the gradient for both the Sector and the $ULA$ model.}
    \label{fig:smpcomp}
\end{figure}

Our algorithm affects the $APL$ and the $CC$ of the network when we use $ULA$ model, (Cf. Fig. \ref{fig:unifbiobeamULA}). On the other hand, it does not affect $|P|$ and $|C|$. No dependency of the $ULA$ model on $|P|$ and $|C|$ is rightly justified because these sets are built when the network was omnidirectional, (Cf. Fig. \ref{subfig:pnodeULA}, \ref{subfig:cnodeULA}). However, there is a reduction of almost $60\%$ and $68\%$ in the $APL$ for higher gradient value and for low gradient value respectively. On the other hand, there is no considerable reduction in the $CC$. The reduction in the $CC$ is only between $19\%$ to $22\%$. Due to variation in $B_{w}$ for different $B_{b}$ in $ULA$ model (Cf. Fig. \ref{fig:gainpatern}), the values obtained for the $APL$, the $CC$ and connectivity are different from that of the Sector model. From the Fig. \ref{fig:smpcomp} we observe that, for higher density networks, the change in the $APL$ for the $ULA$ model is more than that of the Sector model while the $CC$ changes at a much lower rate.

Until now, we have shown that small world properties are achieved and connectivity be increased in a non-uniformly deployed network. However, it is also important to show the complexity of the algorithm. Due to the storage of three required data values in the region formation phase, neighborhood information and the knowledge about being the peripheral node for both itself and its neighbors is needed. Thus the required memory size is of the order \emph{O(3(d+r)+d+1)} where $d$ is the size of the neighborhood and $r$ is the size of reachable centroid nodes. For high-density network, reaching consensus in the region formation and the centroid finding phase is time consuming. However, for a low-density network, the algorithm reaches this consensus quickly.

\section{Useful concepts and Related Work}\label{sec:sec2}

In this section, we define useful concepts giving an overview of the related work. We first define small world concepts in the section \ref{sub:subswn} which form the basis of our research. The need of having long range links for achieving small world properties lead us to discuss beamforming in the section \ref{sub:subamb}. We then define Lateral Inhibition in the section \ref{sub:subli} and Flocking in the section \ref{sub:subflocking}. The definitions of centrality concepts are discussed in the section \ref{sub:subcentral}. Further, we discuss non-uniform deployment in the section \ref{sub:subnonunifdistrib}.

\subsection{Small World Network}\label{sub:subswn}

Inspired by Stanley Milgram's \cite{Milgram} experiment of ``six degrees of separation", Watts et al \cite{Watts} suggested a model for the creation of small world network. Watts et al in \cite{Watts,Wattsbook} showed that rewiring edges of a regular network with a probability $p_r$ results into reduction in the $APL$ of the network while there is very little change in the $CC$. Starting by choosing a random vertex and one of its edge to the vertex's 1 hop neighbor with $p_r$, Watts et al reconnected the edge to a random vertex in the remaining network. Watts et al then considered all other vertices for rewiring. The process of rewiring continued with the edges now connecting the two hop neighbors. This process continued until all the edges were considered. $p_r$ highly affected the rewiring process. Probability $p_r=0$ meant that no rewiring while $p_r=1$ meant complete rewiring of the graph. Using $p_{r}=1$ resulted into complete randomness in the network.

The small world model motivated many research studies, \cite{Helmy,Barabasi,AlbertBarabasi,BarabasiAlbert}, and many models were proposed. Newman,\cite{NewmanReview,Newman}, compiled a comprehensive list of the models on small world. Mostly, the researchers studied two kinds of network structures, one without network growth while another with the network growth. Researchers analyzed the scaling and performance issues for the growing networks \cite{Barabasi,BarabasiAlbert}. Barabasi et al in \cite{AlbertBarabasi,BarabasiAlbert} showed that small world properties also exists in a growing network and there is a \emph{preferential attachment} of the nodes giving rise to ``rich gets richer" property. Barabasi et al showed that the real world networks possess these properties. This led to the behavioral analysis of the networks. On the contrary, assuming spatial wireless ad hoc network without growth, Helmy \cite{Helmy} performed the small world analysis and showed that rewiring of links does not change the structure of the network. Two other results shown in \cite{Helmy} are significant in the context of this paper. First, the $APL$ is reduced at a greater rate when shortcuts are 25\% to 40\% in length of the network diameter. Second, the rate of the $APL$ reduction is more when there are only 0.2\% to 2\% shortcut links. The reduction rate stabilizes when there are more than 2\% shortcut links.

\subsection{Antenna Model and Beamforming}\label{sub:subamb}

Authors of \cite{Bettstetter,Balanis} provided an extensive study of antenna models and defined antenna gain using radiation intensity $u(\theta,\phi)$ where angle $\theta$ is angle with the $z$-axis and $\phi$ with the $xy$-plane as

\begin{equation}\label{eq:1}
    g(\theta ,\phi)=\frac{u(\theta ,\phi )}{\frac{1}{4\pi} \int_{0}^{2\pi }\int_{0}^{\pi }u(\theta ,\phi ) sin\theta d\theta d\phi}
\end{equation}

\begin{figure}
    \centering
    \includegraphics[width=0.3\textwidth]{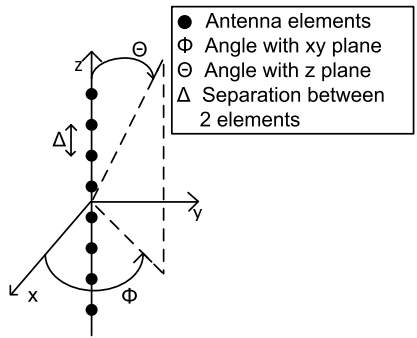}
    \caption{Source: \cite{Bettstetter}, Arrangement of $m=8$ antenna elements in $ULA$ model.}
    \label{fig:figULAUCA}
\end{figure}

Considering $m$ antenna elements and isotropic radiators with same phase shift between them, researchers defined two basic antenna models Uniform Linear Array antenna model ($ULA$), (Cf. Fig. \ref{fig:figULAUCA}), and Uniform Circular Array antenna model ($UCA$). When $m=1$, there is no superimposition of the radiation. This leads to a beam with omnidirectional characteristics. However, when $m>1$, there is a constructive and destructive superimposition of the radiation due to the phase shift between the antenna elements. This leads to a beam with directional characteristics.

\begin{figure}
    \centering
    \mbox
    {
        \hspace{-1cm}
        \subfigure[$B_{b}=0^o, 180^o$]
        {
            \includegraphics[width = 0.3\textwidth]{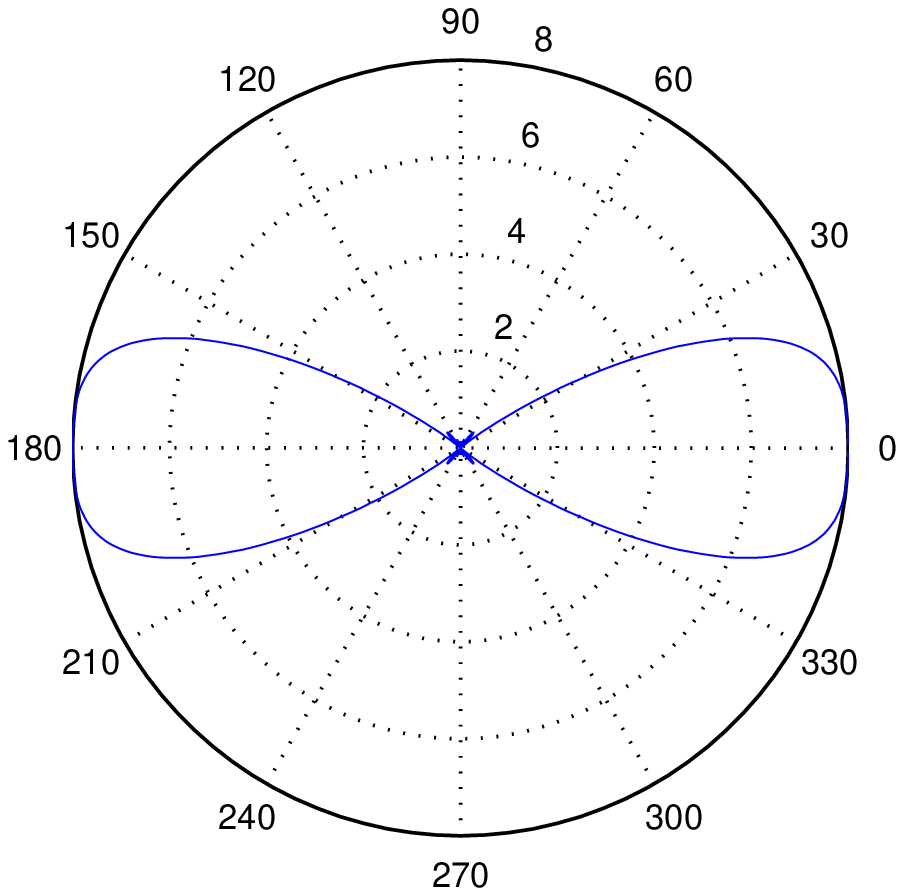}
            \label{subfig:zero}
        }
        \hspace{-1cm}
        \subfigure[$B_{b}=\pm 30^o$]
        {
            \includegraphics[width = 0.3\textwidth]{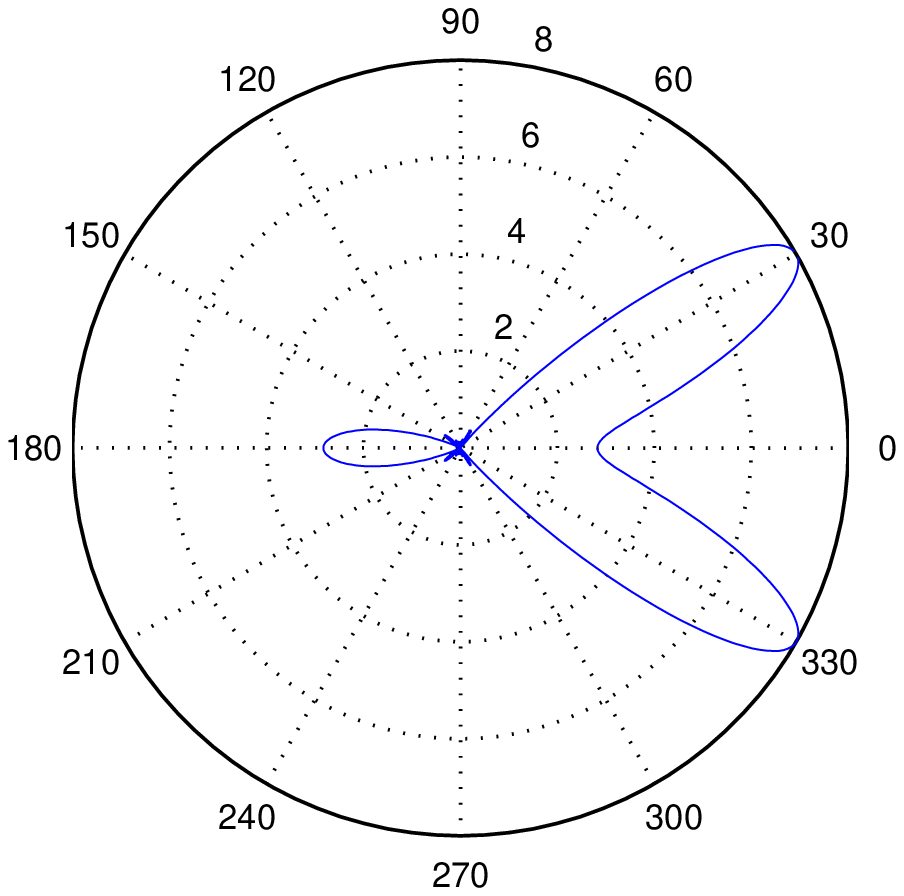}
            \label{subfig:thirty}
        }
    }
    \mbox
    {
        \hspace{-1cm}
        \subfigure[$B_{b}=\pm 60^o$]
        {
            \includegraphics[width = 0.3\textwidth]{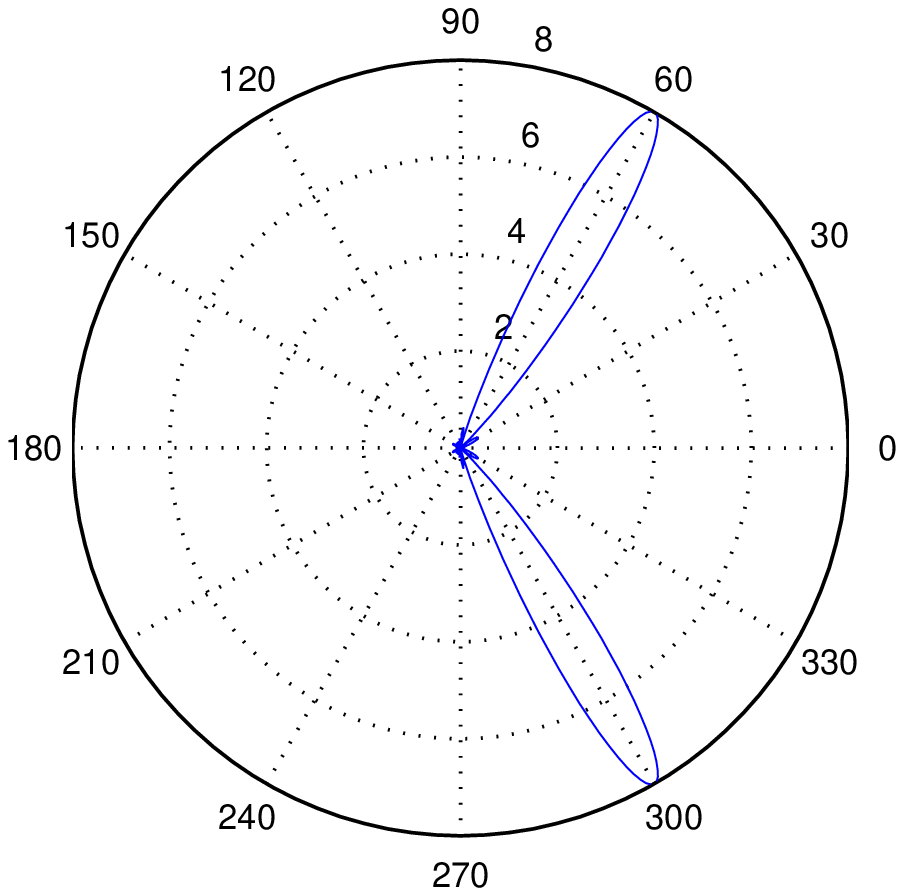}
            \label{subfig:sixty}
        }
        \hspace{-1cm}
        \subfigure[$B_{b}=\pm 90^o$.]
        {
            \includegraphics[width = 0.3\textwidth]{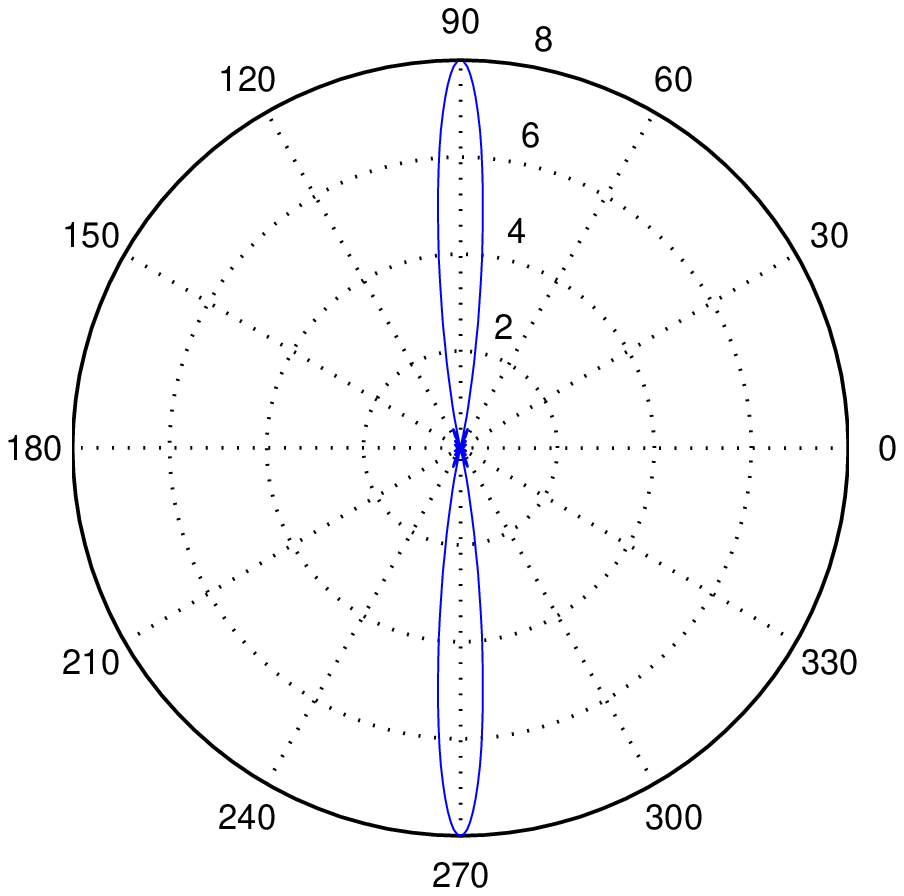}
            \label{subfig:ninty}
        }
    }
    \caption{Source: \cite{Bettstetter}, Gain pattern obtained for different $B_{b}$ and $m=8$ in the $ULA$ model} \label{fig:gainpatern}
\end{figure}

The gain pattern for the $ULA$ antenna model is only dependent on the number of antenna elements. It has no dependency on the boresight direction ($B_{b}$, the direction of maximum radiation intensity, Cf. Fig. \ref{fig:gainpatern}). On the other hand, for the $UCA$ antenna model, gain pattern is dependent on both the number of antenna elements and $B_{b}$.

However, in wireless ad hoc networks, beamforming using $UCA$ model has been well studied. Classical beamforming techniques using $UCA$ model include Random Direction Beamforming ($RDB$) \cite{Bettstetter,Vilzmann,VilzmannBettstetterHartmann} and beamforming based on the angle of incidence and packet flow. Bettstetter et al \cite{Bettstetter} studied the use of $RDB$ with the path probability to improve the connectivity in the wireless networks. Vilzmann et al \cite{VilzmannWidmer} derived low complexity techniques for beamforming and proposed Maximum Node Degree Beamforming ($MNDB$). In $MNDB$ the nodes directed their beams towards the node that had maximum degree. The authors found that $MNDB$ leads to less number of inter-cluster connections but had more intra-cluster connections. To overcome this drawback, the authors proposed Two-hop Node Degree Beamforming ($TNDB$). In $TNDB$ the nodes directed their beams towards the node that had maximum two-hop neighborhood. The authors showed that $TNDB$ outperforms both $RDB$ and $MNDB$. Other works on beamforming include \cite{VilzmannWidmer,Kiese,Yu,Li}. However, most of these studies were concentrated on nodes that were uniformly distributed at random in the given area but very few among them talk about non-uniform distribution of the nodes. Considering all nodes use directional beams, \cite{Bettstetter,Vilzmann,VilzmannWidmer,Kiese,Yu,Li} addressed connectivity very well but do not discuss the impact on the $APL$ and the $CC$. Table \ref{table:table1} illustrates a comparison between these studies. On the other hand, studies related to the small world properties lack connectivity analysis for the non-uniformly distributed network. Table \ref{table:table2} illustrates comparisons between various studies performed in the direction of achieving small world properties in wireless ad hoc networks and our model.

\begin{table*}
    \centering
    \begin{tabular}{|l|l|l|l|l|l|}
        \hline
        \textbf{Parameter\textbackslash} & \textbf{Vilzmann} & \textbf{Widmer} & \textbf{Kiese} & \textbf{Yu} & \textbf{Li} \\
        \textbf{Reference} & \textbf{et al \cite{Vilzmann}} & \textbf{et al \cite{VilzmannWidmer}} & \textbf{et al \cite{Kiese}} & \textbf{et al \cite{Yu}} & \textbf{et al \cite{Li}} \\
        \hline
        Transmission mode & Directional & Directional & Directional & Directional & Both\\
        \cline{1-6}
        Reception mode & Directional & Directional & Directional & Omnidirectional & Both\\
        \cline{1-6}
        Mobility & No & Yes & No & No & No\\
        \cline{1-6}
        Beam width & Depends on & Constant & Constant & Optional & Constant,\\
        & beam direction & & & & switched beam\\
        & & & & & antenna\\
        \cline{1-6}
        Beam direction & Random & Optional & Optional & Optional & Random\\
        \cline{1-6}
        Antenna model & $UCA$ & $UCA$ & $UCA$ modeled & Sector & Keyhole\\
        & & & as keyhole & &\\
        \cline{1-6}
        Node distribution & Uniform & Uniform and & Non-Uniform & Not specified & Uniform\\
        & & Non-Uniform & & &\\
        \hline
    \end{tabular}
    \caption{Comparison between various studies in the direction of beamforming that focus on connectivity. Due to extensive literature, we only consider a limited set of research studies here.}
    \label{table:table1}
\end{table*}

\begin{table*}
    \centering
    \begin{tabular}{|l|l|l|l|l|l|l|}
        \hline
        \textbf{Parameter\textbackslash} & \textbf{Our Model} & \textbf{Banerjee} &\textbf{Guidoni} & \textbf{Helmy} & \textbf{Sharma} & \textbf{Verma}\\
        \textbf{Reference} & & \textbf{et al \cite{Banerjee}} &\textbf{et al \cite{GuidoniLoureiro}} & \textbf{et al \cite{Helmy}} & \textbf{et al \cite{Sharma}} & \textbf{et al \cite{Verma}}\\
        \hline
        Shortcut Creation & Rewiring & Rewiring &Addition & Addition & Addition & Addition\\
        \cline{1-7}
        Node distribution & Non-Uniform & Uniform & Uniform & Uniform & Uniform & Uniform \\
        \cline{1-7}
        External & No & No & High range & - & Wired & Two radios \\
        infrastructure& & & Sensor & & & for each node\\
        \cline{1-7}
        Global knowledge & No & No & Yes & Yes & Yes & Yes\\
        \cline{1-7}
        Density of nodes & Low & High & High & High & - & Low\\
        \cline{1-7}
        Shortcut Edge & Directed & Directed & Undirected & Undirected & Undirected &Directed\\
        \cline{1-7}
        Shortcut direction & Towards & Longest & Random, & Random & Random & Random\\
        & centroid of & Traffic Flow & towards sink & & &\\
        & other region &path & & & &\\
        \cline{1-7}
        Shortcut length & Function of & Function of & Constant & Limited & Constant & Constant\\
        & antenna& node density & & & &\\
        & elements& & & & &\\
        \cline{1-7}
        Shortcut width & Depends on & Depends on & Constant & - & - & Constant\\
        & Shortcut Length & Shortcut Length& & & & \\
        \cline{1-7}
        Prob. of Shortcut & $(0,1]$ based & Based on & $\in (0,1]$ & $\in (0,1]$ & function of & $\in (0,1]$\\
        creation &on model & centrality & & & network size & \\
        &parameters & values& & & &\\
        \cline{1-7}
        \hline
        \multirow{3}{*}{Performance metric} & Path length, & Path length,& Path length, & Path length, & Path length, & Path length,\\
        & Clust. Coeff. & Connectivity & Clust. Coeff. & Clust. Coeff. & Energy & Clust. Coeff.,\\
        & Connectivity & & & & & degree\\
        \hline
    \end{tabular}
    \caption{Comparison between various research studies in the direction of achieving small world properties in the wireless ad hoc networks. Other research studies in this direction consider use of external infrastructure of at least two radios.}
    \label{table:table2}
\end{table*}

\subsection{Lateral Inhibition}\label{sub:subli}

Lateral Inhibition is a process by which cells of animal tissues, based on the properties of neighbor cells, decide whether to perform a task or not. Lateral Inhibition ensures that the cells that perform the tasks are equidistant from each other. This helps in producing regular patterns throughout the surface. Lawrence \cite{Lawrence} modeled Lateral Inhibition as, when a cell performs a task, it inhibits its neighbors within $h$ hops from performing that task thereby resulting into equally spaced uninhibited cells. Lateral Inhibition thus creates clusters where the cluster heads are uninhibited nodes distributed over an area. Nagpal et al \cite{Nagpal,NagpalMamei} described a simple algorithm to achieve Lateral Inhibition. In the algorithm, the cells assign themselves a random number. Each cell starts to count backwards. If before reaching $0$, a node receives an inhibition signal from the neighboring cell, the cell stops counting otherwise sends out an inhibition signal to all its neighbors. Nagpal et al \cite{Nagpal,NagpalMamei} showed that the $hopcount$ used to create the cluster greatly affects the number of clusters formed.

Recent studies revealed that Lateral Inhibition can be achieved in an optimal way \cite{Afek}. Inspired by the tissue of the fruit fly, Afek et al \cite{Afek} modeled distributed Lateral Inhibition using local information and requiring only two exchange mechanisms. These exchange mechanisms are, first, broadcasting a single control bit to the neighbors with certain probability and second, if the node receives no message from the neighbors, it sends out a control bit to inhibit its neighbors. As a variation to Nagpal et al's algorithm, the algorithm used a probabilistic approach that varied over time in an increasing manner to perform Lateral Inhibition. The runtime complexity of the algorithm was of the order $O(\log^2|V|)$ where $|V|$ was the number of nodes in the system. Due to single bit exchange messages over single hop, the algorithm had a low message complexity.

\subsection{Flocking}\label{sub:subflocking}

\begin{figure*}
    \centering
    \mbox
    {
        \subfigure[Alignment]
        {
            \includegraphics[width = 0.25\textwidth]{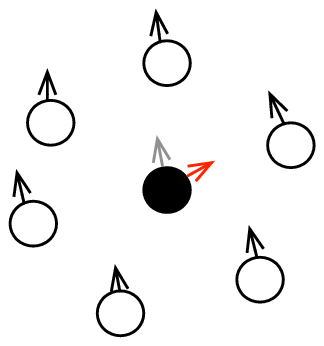}
            \label{subfig:alignment}
        }
        \hspace{-0.5cm}
        \subfigure[Separation]
        {
            \raisebox{0.5cm}{
            \includegraphics[width = 36mm]{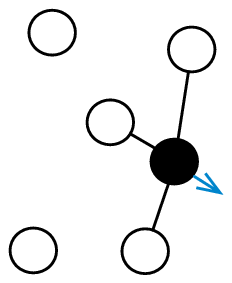}
            \label{subfig:separation}}
        }
        \hspace{-0.7cm}
        \subfigure[Cohesion]
        {
         \raisebox{0.5cm}{
            \includegraphics[width = 45mm]{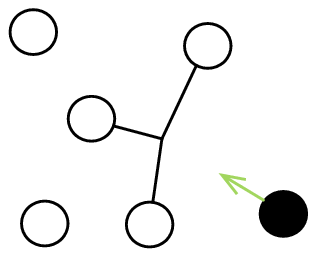}
            \label{subfig:cohesion}}
        }
        \hspace{-0.5cm}
        \subfigure
        {
            \raisebox{3cm}
            {
                \includegraphics[width = 0.25\textwidth]{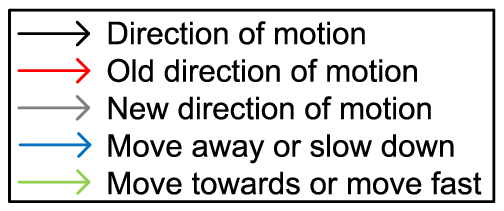}
            }
            \label{subfig:figflocklegend}
        }
    }
    \caption{Source: \cite{Reynolds}, Depiction of three Flocking rules.} \label{fig:figflocking}
\end{figure*}

Flocking, \cite{Reynolds}, was first modeled by Reynolds in order to simulate the birds' behavior. In nature, flocking is observed in many other social living organisms like cattle, fishes and humans. Reynolds, while modeling Flocking, termed each social entity as a \emph{boid} and formulated three very simple rules, (a) Alignment (b) Separation and (c) Cohesion. Reynolds defined Alignment rule as the direction matching of a \emph{boid} with its neighbors. He defined Separation rule as the collision avoidance with neighborhood \emph{boids} and Cohesion rule as the tendency of a \emph{boid} to remain as close to its neighbors as possible and not stray. The Fig. \ref{fig:figflocking}(a), shows that the \emph{boid} orients itself in the direction in which its neighbors were moving. The Fig. \ref{fig:figflocking}(b), shows that the \emph{boid} has to move away from the neighbors in order to avoid collision while the Fig. \ref{fig:figflocking}(c) shows that the \emph{boid} moves towards the centroid of the neighbors in order to remain close to its neighborhood. Couzin in \cite{Couzin} formulated mathematical explanation of these rules. Due to the motion of a \emph{boid}, velocity and displacement were associated with the \emph{boid}. Alignment rule was modeled using the direction of a \emph{boid} while Separation and Cohesion were modeled using both velocity and the displacement.

Recent studies have revealed the use of Flocking in solving various problems in wireless ad hoc networks. Antoniou et al \cite{Antoniou} used Flocking to provide efficient congestion control mechanism by computing the congestion at the neighbor nodes while \cite{Kadrovach} used the Separation rule for the efficient placement of nodes to maximize the coverage area.

\subsection{Centrality}\label{sub:subcentral}

Decades of research on network and graph theory has led researchers to derive many fundamental concepts related to the importance of a node in the network. The concept of centrality was one such concept that was developed and used to address the topological characteristics of the network nodes. Proposed definitions of centrality measures include those that use global parameters as well as those that only use local information. Some examples of global centrality measures are Socio-Centric Betweenness \cite{Freeman,Freemanlc} and Closeness Centrality \cite{Freeman} while Degree Centrality \cite{Freeman} and Egocentric Betweenness Centrality \cite{Everett,Daly} are examples of the local centrality measure.

\subsubsection{Socio-Centric Betweenness Centrality}\label{subsubsec:scbc}

The Socio-Centric Betweenness Centrality, \cite{Freeman,Freemanlc}, is the measure of the number of shortest paths passing through the node thereby expressing the most important node in the network and through which most of the communication takes place. The Socio-Centric Betweenness is a frequency measure and requires the global network knowledge. Usually nodes with high degree and those that are acting as the bridge nodes tend to have relatively high Socio-Centric Betweenness. Mathematically the Socio-Centric Betweenness of a node $v$ is

\begin{equation}
    BC_v=\sum_{}\frac{sp(v)}{sp}
\end{equation}

where $sp(v)$ is the number of shortest paths between any two nodes that pass through $v$ while $sp$ is the total number of shortest paths in the network.

\subsubsection{Egocentric Betweenness Centrality}\label{subsubsec:ebc}

Aiming to compute the Betweenness centrality using local properties, \cite{Everett,Daly} proposed the Egocentric Betweenness Centrality measure. Everett in \cite{Everett} computed the Egocentric Betweenness using upper diagonal adjacency matrix $A_{v}$. $A_{v}$ is created considering 1 hop neighborhood of the node $v$. Consider $I$ to be the identity matrix, then the sum of the inverse of all non-zero elements in $A_{v}^{2}$ along $[I-A_{v}]$ is the Egocentric Betweenness of the node.

Marsden in \cite{Marsden} performed an empirical study to find the relation between the two types of Betweenness, the Socio-Centric and the Egocentric Betweenness, and found that the Egocentric Betweenness is strongly correlated to the Socio-Centric Betweenness and it can be used when global network information is lacking.

\subsubsection{Closeness Centrality}\label{subsubsec:closenessc}

The Closeness Centrality \cite{Freeman} on the other hand is the measure of how fast a node can transfer data to all the nodes. The Closeness Centrality is the fraction of shortest distance between a node to all other nodes in the network. Assuming $sd(v,w)$ be the shortest distance between node $v$ and $w$, the Closeness Centrality of $v$ is

\begin{equation}
    C_{v}= \frac{1}{\sum_{w\neq v,w\in V} {sd(v,w)}}
\end{equation}

A node with the highest Closeness Centrality value is the centroid of the network.

As all the centrality measures convey different information, it is not necessary that a node having high value for one centrality measure also have high values for the others. Many other types of centralities, such as, Bridging Centrality, Eigen Vector Centrality and Spectral Centrality also exist. We refrain ourselves from describing them in detail. However, Katsaros, \cite{Katsaros}, provided a brief survey on these centrality measures.

\subsection{Non-Uniform distribution of nodes}\label{sub:subnonunifdistrib}

\begin{figure*}[ht]
    \centering
    \mbox
    {
        \subfigure[Uniform node distribution with the node density.]
        {
            \includegraphics[width = 0.45\textwidth]{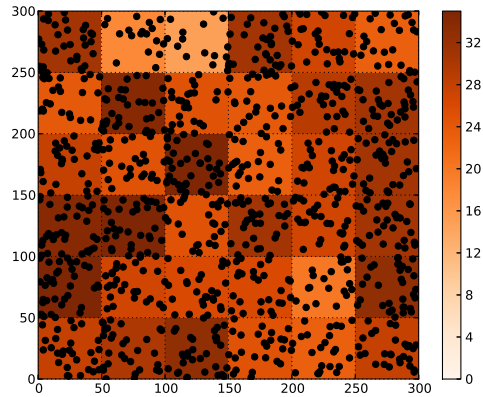}
            \label{subfig:figVincentnodedistbunif}
        }\quad
    }
    \mbox
    {
        \subfigure[Distribution after applying the Thinning process with $r_b=15$ and $\ell_{min}=6$ on the distribution as in the Fig. \ref{subfig:figVincentnodedistbunif}.]
        {
            \includegraphics[width = 0.45\textwidth]{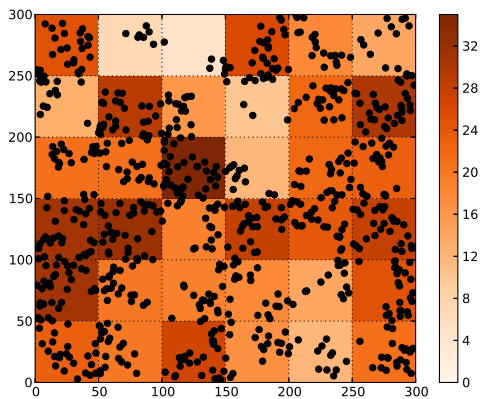}
            \label{subfig:figVincentnodedistbnonunif}
        }
        \hspace{1cm}
        \subfigure[Distribution pattern using Kernel method for the Fig. \ref{subfig:figVincentnodedistbnonunif}.]
        {
            \includegraphics[width = 0.45\textwidth]{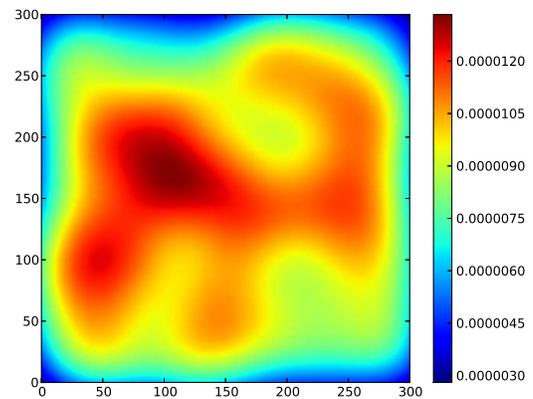}
            \label{subfig:figVincentdensity}
        }
    }
    \caption{Node Distribution.}
    \label{fig:distribution}
\end{figure*}

Many non-uniform deployment strategies have been proposed, \cite{Weijen,LeBoudec,Hu,Aitsaadi,Riihijarvi,BettstetterGyarmati}. We take insights from Bettstetter et al, \cite{BettstetterGyarmati}, node deployment strategy. Bettstetter et al proposed the use of thinning process to generate a non-uniform node deployment. The authors started with uniform distribution of nodes in a given region, then pruned the nodes based on two factors, transmission radius, $r_b$, and the number of neighbor nodes, $\ell_{min}$. If the node had at least $\ell_{min}$ neighbors within $r_b$, the node was not removed else it was removed. Schilcher et al, \cite{SchilcherGyarmati}, formulated and measured the degree of non-uniformity of this pruned network. Schilcher et al divided the region into smaller sub-regions and estimated the number of nodes in the sub-region. The estimated value was then used to calculate the non-uniformity index, $hIndex$. The Fig. \ref{subfig:figVincentnodedistbnonunif} shows the deployment achieved when the thinning process is applied to the deployment shown by the Fig. \ref{subfig:figVincentnodedistbunif}. The Fig. \ref{subfig:figVincentdensity} shows the density distribution of nodes using kernel method.

\section{Future Work}\label{sec:sec9}

A Number of extensions to our algorithm can be visualized. Identifying the optimal gradient size to choose for the determination of minimal peripheral set of nodes is one way of extending our work. We are currently working on how we can apply game theory to successfully find the minimal peripheral set. We believe that by applying game theory nodes can determine what the suitable gradient size is and can reduce asymmetric links further.

We would also like to extend our algorithm to support dynamic environment and asynchronous operation. Dynamic environments are likely to result in frequent changes to the state of the node. Any change in the state of the node would require reconfiguring in the network using the proposed algorithm. Information available at the neighborhood nodes would be helpful in learning about the previous configuration. This learning could be \emph{docitive} \cite{Giupponi}, meaning, partial learning from the neighborhood states could make nodes infer about the previous good configuration so that reconfiguration can be done easily and quickly. This will also help us to address the unaddressed paradigms of \cite{Prehofer}. Further, we would like to address network lifetime of the network when implementing our algorithm.

\section{Conclusion}\label{sec:sec8}

In this paper, we have presented an algorithm for achieving small world properties using beamforming and bio-inspired techniques in a wireless ad hoc network. Our algorithm works using locally available information and does not require the knowledge of the network. We have also removed the possibility of requirement of any external infrastructure for achieving our goal. Through our algorithm, we have shown how isolated communities can collaborate and connect with each other to achieve better and faster communication. Bio-Inspired techniques like Lateral Inhibition helped us to form communities within the network for the reduced message complexity while the Flocking analogy helped us to determine beam properties. Our results show that for both theoretical and realistic antenna models and relatively high-density networks, there is a reduction in the $APL$ by almost $40\%$ to $68\%$ for $g\in[3,10]$. On the other hand, reduction in the $CC$ is between $19\%$ to $38\%$. Our results also show improvement in the connectivity. The increase in the size of the $GSCC$ for the non-uniformly distributed directional network is around $10\%$ for high density network while it is around $61\%$ for relatively low density networks.

\bibliography{biblio}
\bibliographystyle{IEEEtr}

\end{document}